\documentclass[10 pt,a4paper,twoside]{article}

\usepackage[utf8]{inputenc}
\usepackage[english]{babel}
\usepackage[left=2cm,right=2cm,top=2cm,bottom=2cm]{geometry}

\usepackage{amsmath}
\usepackage{amssymb}	
\usepackage{amsfonts}
\usepackage{mathtools}
\usepackage{bigints}
\usepackage[linktocpage, hyperfootnotes=false]{hyperref}

\usepackage{cite}

\usepackage{graphicx}
\usepackage{float}
\usepackage{caption}
\usepackage{subcaption}
\graphicspath{ {Images/} }
\usepackage{wrapfig}
\usepackage[none]{hyphenat}

\usepackage{multicol}

\usepackage{authblk}
\author[1,2\footnote{anujmishra.physics@gmail.com}]{\small Anuj Mishra}
\author[2\footnote{schakraborty.math@gmail.com }]{\small Subenoy Chakraborty}

\affil[1]{\footnotesize \textit{National Institute of Technology, Rourkela, Odisha, 769008, India.}}
\affil[2]{\footnotesize \textit{Department of Mathematics, Jadavpur University, Kolkata-700032, India.}}

\date{}
\begin{document}

\title{On the trajectories of null and timelike geodesics in different wormhole geometries}

\maketitle

\begin{abstract}
The paper deals with an extensive study of null and timelike geodesics in the background of wormhole geometries. Starting with a spherically symmetric spacetime, null geodesics are analyzed for the Morris-Thorne wormhole(WH) and photon spheres are examined in WH geometries. Both bounded and unbounded orbits are discussed for timelike geodesics. A similar analysis has been done for trajectories in a dynamic spherically symmetric WH and for a rotating WH. Finally, the invariant angle method of Rindler and Ishak has been used to calculate the angle between radial and tangential vectors at any point on the photon's trajectory.	
\end{abstract}

\section{Introduction}
In general relativity, a wormhole (WH) is considered to be a tunnel through which two distant regions of spacetime can be connected \cite{visser}. Long back in 1916, Flamm \cite{flamm} introduced the idea of wormhole, analyzing at that time the recently discovered Schwarzschild solutions. In 1935, Einstein and Rosen \cite{einstein} constructed WH type solution considering an elementary particle model as a bridge connecting two identical sheets. This mathematical representation of space being connected by a WH type solution is known as ``Einstein-Rosen bridge". Wheeler \cite{jwheeler,jjwheeler} in the 1950s considered WHs as objects of quantum foam connecting different regions of spacetime and operating at the Planck scale. Subsequently, using this idea, Hawking \cite{hawking} and collaborators introduced the idea of Euclidean wormholes. But these types of WHs are not traversable and, in principle, would develop some type of singularity \cite{geroch}. However, these hypothetical shortcut paths, i.e., traversable WHs, have been rekindled by the pioneering work of Morris and Thorne \cite{mt} which is considered as the modern renaissance of WH physics. Subsequently, it was claimed that there is no strong ground \cite{harrison,novikov} for the energy conditions and hence one considered WH,  with two mouths and a throat, to be an object of nature, i.e., an astrophysical object.

On the other hand, in general relativity, WH physics is a specific example where the matter stress-energy tensor components are evaluated from the spacetime geometry by solving Einstein's field equations. But for a traversable WH, the stress-energy tensor components so obtained always violate the null energy condition \cite{visser, mt}. As the null energy condition (NEC) is the weakest of all the classical energy conditions, its violation signals that the other energy conditions are also violated. In fact,  they violate all the known pointwise energy conditions and averaged energy conditions, which are fundamental to the singularity theorems and theorems of classical black hole thermodynamics. Generally, it is believed that a classical matter obeys energy conditions \cite{ellis} but, in fact, it is known that they also get violated by some quantum fields (namely as regards the Casimir effect and Hawking evaporation \cite{scalar}). Further, for a quantum system in classical gravity, it is found that the averaged weak or null energy condition(ANEC), which states that the integral of the
energy density as measured by a geodesic observer is non-negative, could also be violated by a small amount \cite{tipler,roman}.

Finally, it is worth to mention a few important dynamical WH solutions. Hochberg and Visser \cite{Hochberg} and Hayward \cite{Hayward}  independently formulated the dynamical WH solutions, choosing a quasi local definition of the WH throat in a dynamical spacetime. Accordingly, WH throat is a trapping horizon \cite{hhayward} of different kind but again matter in both of them violates the NEC. On the other hand, Maeda,et al. \cite{maeda} have developed another class of dynamical WHs (cosmological WHs) which are asymptotically FRW spacetimes with big bang singularity at the beginning. This class of WHs contain matter which not only obey NEC but also the dominant energy condition everywhere. These two classes of dynamical WHs are distinct from the geometrical point of view. For the former one, the WH throat is a 2D surface of non-vanishing minimal area of a null hypersurface, while for the later one, there is no past null infinity due to the initial singularity. Hence, the WH throat is defined only on a space-like hypersurface and the spacetime is trapped everywhere without any trapping horizon \cite{mmaeda}. Recently, Lobo et al. \cite{lobop,loboq,lobor} formulated wormhole solutions which are dynamically generated using a single charged fluid. Also, dynamical WHs are considered with a two-fluid system \cite{campo,ccampo}, for a matter distribution relevant to present day observations \cite{ssc} and using the mechanism of particle creation \cite{sc}. Then for evolving WH\footnote{These are not as popular as static WHs and also not well understood}, one may refer to Refs.\cite{lobo,teo,xxx12,xxx13,mella}. 

The particle motion in wormhole spacetimes is an important issue related to traversable WHs. It is interesting to examine whether a timelike or null geodesic can tunnel through the throat of the WH. Cataldo \textit{et al.}\cite{reva} studied motion of test particles in the background of zero tidal force Schwarzschild-like WH spacetime. They showed that particles moving along the radial geodesics reach the throat with zero tidal velocity in finite time while the particle velocity reaches maximum at infinity if it travels along a radially outward geodesic. For non-radial geodesics on the other hand, the particles may cross the throat with some restrictions. Olmo \textit{et al.}\cite{revb} carried out a detailed investigation of the geodesic structure for three possible WH configurations, namely: Reissner–Nordstr$\ddot{o}$m-like WH, Schwarzschild-like WH and Minkowski-like WH. They have shown that it is possible to have geodesically complete paths for all these WH spacetimes. Culetu\cite{revc} examined both timelike and null geodesics for a WH belonging to the Planck world ( WHs whose throat size is of the order of the Planck length $l_P$) where quantum fluctuations are supposed to exist and the spacetime smoothness seems to break down. Muller\cite{revd} also studied null and timelike geodesics in WH configuration using elliptic and Jacobian integral functions. He showed that it is possible to connect two distant events geodesically. Regarding geodesic study in non-static WHs, recently Chakraborty and Pradhan\cite{reve} have studied the geodesic structure of the rotating traversable Teo WH. Also, Nedkova \textit{et al.}\cite{revf} discussed the shadow of a class of rotating traversable WH in the framework of general relativity. They showed that the images depend on the angular momentum of the WH and the inclination angle of the observer. Finally, it is worthy to mention the work of Ellis \cite{revg}. He constructed a static, spherically symmetric, geodesically complete, horizonless spacetime manifold with a topological hole (drainhole) at its center by coupling the geometry of  Schwarzschild spacetime to a scalar field. It is found that on one side of the drainhole the manifold is asymptotic to a Schwarzschild manifold with positive mass parameter `m', and on the other to a Schwarzschild manifold with negative mass parameter `$\bar{m}$', with the condition $-\bar{m}>m$. As a consequence, there is attraction of particles on one side while there is repulsion on the other side (with higher strength). 

The present work presents a detailed investigation of both timelike and null geodesics both for static and dynamical WHs. The paper is organized as follows: Sect. \ref{staticwh} deals with static spherical WHs in which null and timelike geodesics are studied in great detail. A similar geodesic analysis is presented for dynamical WH in Sect. \ref{dynamicwh} and rotating WH in Sect. \ref{rotatingwh}. Sect. \ref{invariantangle} uses the invariant angle method of Rindler and Ishak to calculate the angle between radial and tangential vectors at a point on the photon's trajectory. Finally, the paper ends with a short discussion and concluding remarks in Sect. \ref{conclusion}. Throughout our analysis, we have chosen to work with wormholes whose material extends all the way from the throat out to infinity.

\section{Trajectories in a spherically symmetric and static geometry} \label{staticwh}
The metric for a general spherically symmetric and static metric can be written as (Ref.\cite{schutz,mtw}),
\begin{equation} 
ds^2=-A(r)dt^2+B(r)dr^2+C(r)d\Omega ^2 \label{a}
\end{equation}
where, $$\lim_{r\to\infty} A(r) =\lim_{r\to\infty} B(r)=1\ \ \ \text{and,}\ \ \ \lim_{r\to\infty} C(r)=r^2$$
An important relation between momenta one-forms of a freely falling body and the background geometry is given by the geodesic equation\cite{schutz},
\begin{equation}
\frac{dp_\beta}{d\lambda}=\frac{1}{2}g_{\nu \alpha, \beta} p^\nu p^\alpha \label{b}
\end{equation}
where $\lambda$ is some affine parameter. 
This relation tells us immediately that if all the components of $g_{\alpha \nu}$ are independent of $x^\beta$ for some fixed index $\beta$, then $p_{\beta}$ is a constant along any particle's trajectory, i.e., a constant of motion. Now, if we work in the equatorial plane by setting $\theta=\pi/2$, then, in $Eq.$\eqref{a}, all the $g_{\alpha \beta}$ become independent of $t,\theta, \phi$ (cyclic coordinates). That means that we can find the respective Killing vector fields $\delta^\nu_\alpha \partial_\nu$ with $\alpha$ as cyclic coordinates. Now, since $p_t$ and $p_\phi$ are constants of the motion, we will set them as
\begin{equation}
p_t=-E \text{,} \ \ \ \ p_\phi= L \label{c} 
\end{equation}
where $E$ is the energy and $L$ is the angular momentum of the photon or a  particle as measured by observers at asymptotically flat regions far from the source. Thus, we get
\begin{equation}
p^t=\dot{t}=g^{t\nu}p_\nu=\frac{E}{A(r)} \text{,} \ \ \ \ \ p^\phi=\dot{\phi}=g^{\phi \nu}p_\nu=\frac{L}{C(r)}\ \ \ \ \ \text{and let,} \ \ \ p^r=\frac{dr}{d\lambda}=\dot{r} \label{d}
\end{equation}
where the dot represents the derivative w.r.t. some affine parameter $\lambda$.

\subsection{Null geodesics}
Now, for null-geodesics, we have $ p_\alpha p^\alpha=0 $. Thus,
\begin{equation}
 \dot{r}^2=\frac{1}{B(r)} \bigg( \frac{E^2}{A(r)}-\frac{L^2}{C(r)}\bigg) \label{e}
\end{equation}
Using $Eq.$\eqref{d} and \eqref{e}, we can write the equation of the photon trajectory in terms of the impact parameter, $\mu=L/E$, as:
\begin{equation} \label{czb}
\boxed{
\bigg(\frac{dr}{d\phi} \bigg)^2= \frac{C^2(r)}{\mu^2 B(r)}\bigg[\frac{1}{A(r)}-\frac{\mu^2}{C(r)}\bigg]}
\end{equation}
If we assume that the geometry is caused by a source of radius $r_s$, then the photon coming from infinity will not hit the surface if there exists a solution $r_o>r_s$ for which $\dot{r}^2=0$. We then call $r_o$ as the distance of closest approach or the turning point. In that case,
\begin{equation} \label{f}
\frac{L^2}{E^2}=\frac{C(r_o)}{A(r_o)} \hspace{0.5cm} \{ \text{if } B^{-1}(r) \neq 0 \text{ for any } r>r_s\} 
\end{equation}
The impact parameter then becomes, 
\begin{equation} \label{g}
 \boxed{ \it{ \mu}=\frac{L}{E}=\pm \sqrt{\frac{C(r_o)}{A(r_o)}}}  
\end{equation}
Using $Eq.$\eqref{czb}, we can write,
\begin{equation}
\frac{d\phi}{dr}= \pm \sqrt{\frac{B(r)}{C(r)\bigg[ \bigg(\frac{A(r_o)}{A(r)} \bigg)\bigg(\frac{C(r)}{C(r_o)} \bigg)-1 \bigg]      }} \label{h}
\end{equation}
Now, if a photon coming from the polar coordinate $\lim_{r\to\infty}(r,-\pi/2-\alpha/2)$ passes through a turning point at $(r_o,0)$ before approaching the point $\lim_{r\to\infty}(r,\pi/2+\alpha/2)$, then this $\alpha$, which is a function of $r_o$, is what we refer to as the deviation/deflection angle, given by (Ref.\cite{ew}
\begin{equation}
\boxed{
\alpha(r_o)=-\pi + 2\int_{r_o}^{\infty} \frac{ \sqrt{B(r)}dr }{ \sqrt{C(r)} \sqrt{ [A(r_o)/A(r)] [C(r)/C(r_o)] -1 } }                
} \label{i}
\end{equation}
However, it is possible that a photon might get trapped in a sphere of constant $r$ and thus may not approach $\lim_{r\to\infty}(r,\pi/2+\alpha/2)$. In that case, the integral will diverge. Such spheres are called photon spheres; they are discussed in sec.\ref{psp}.

\subsection{Morris-Thorne wormhole}
\begin{figure}[t]
\centering
\begin{subfigure}{.5\textwidth}
  \centering
  \includegraphics[scale=0.62]{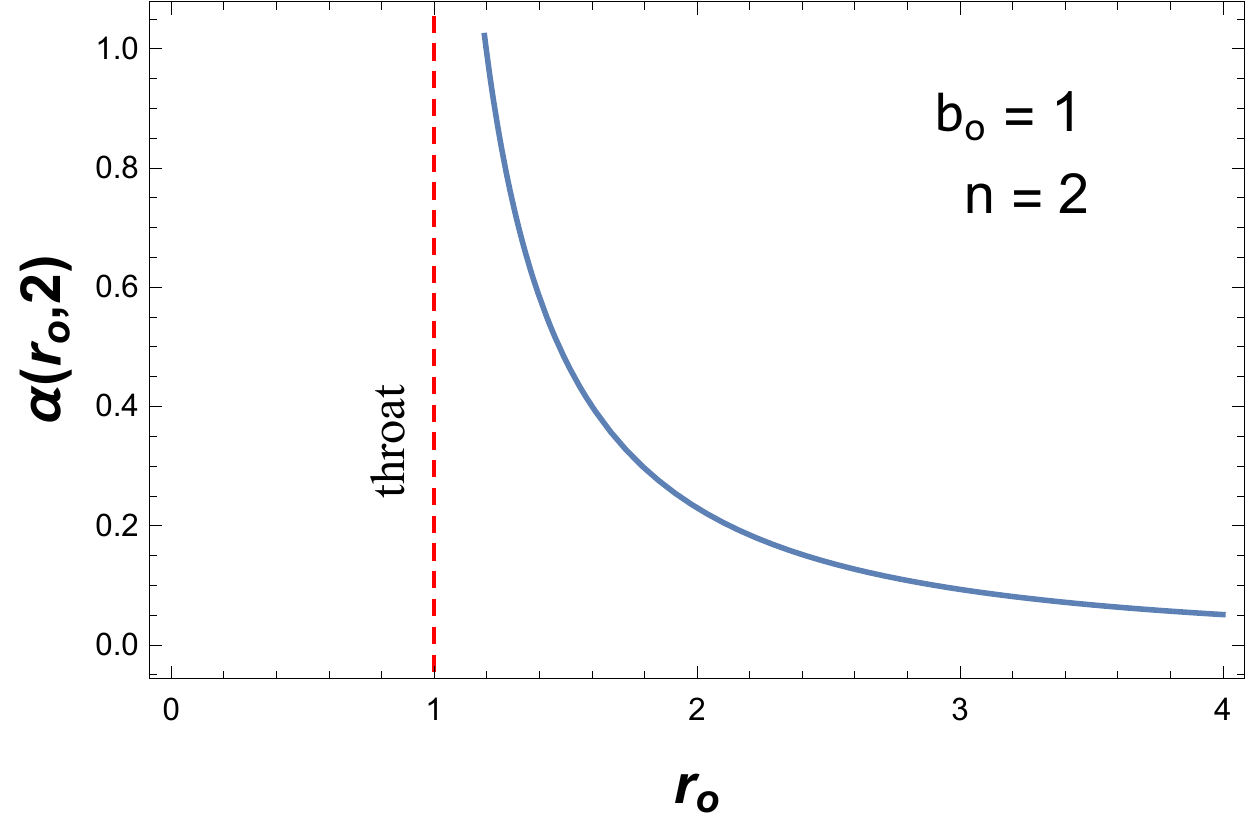}
  \caption{\ $\alpha(r_o,2)$ vs. $r_o$}
  \label{plotnmt}
\end{subfigure}%
\begin{subfigure}{.5\textwidth}
  \centering
  \includegraphics[scale=0.6]{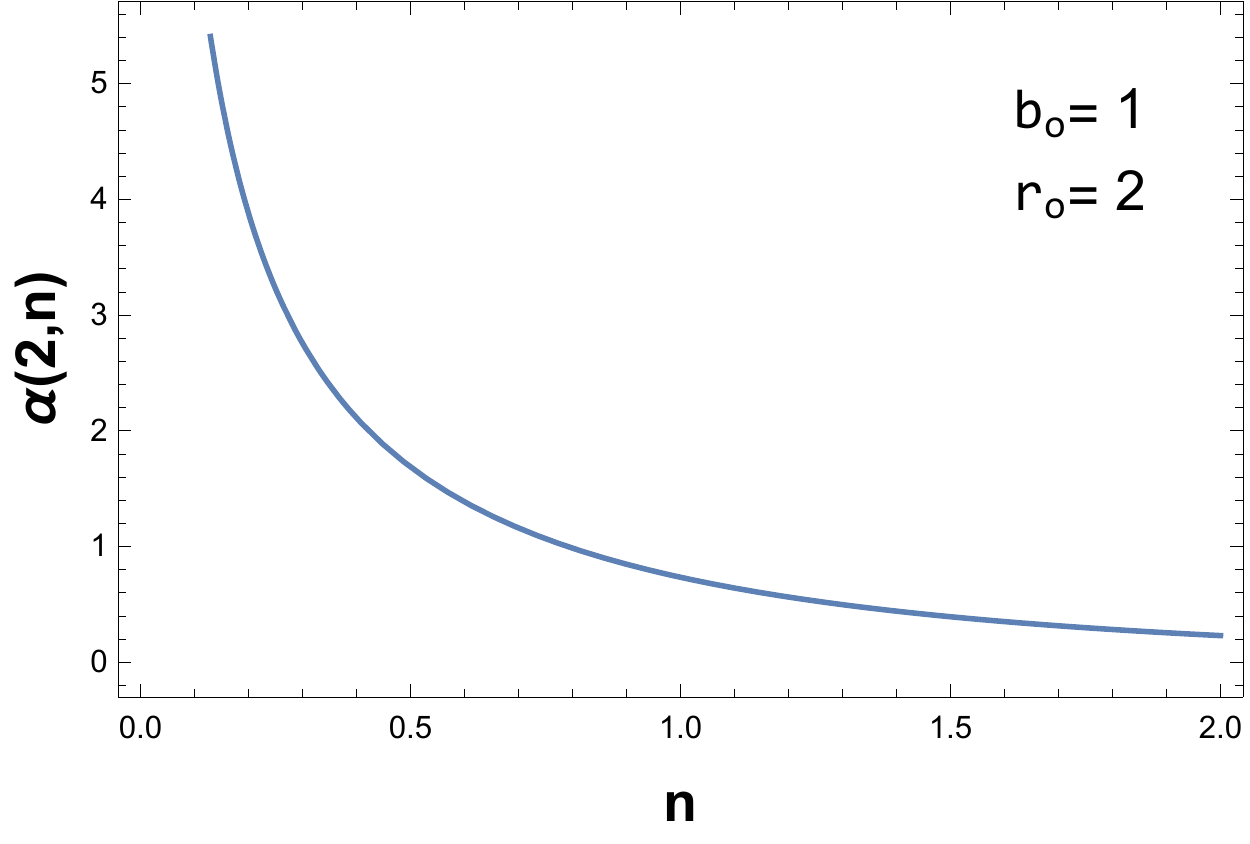}
  \caption{$\alpha(2,n)$ vs. $n$}
\end{subfigure}
\caption{The figures show how $\alpha(r_o,n)$ depends upon its parameters.}
\label{angle}
\end{figure}

\begin{figure}[h]
\begin{subfigure}{.5\textwidth}
\centering
  \includegraphics[scale=0.43]{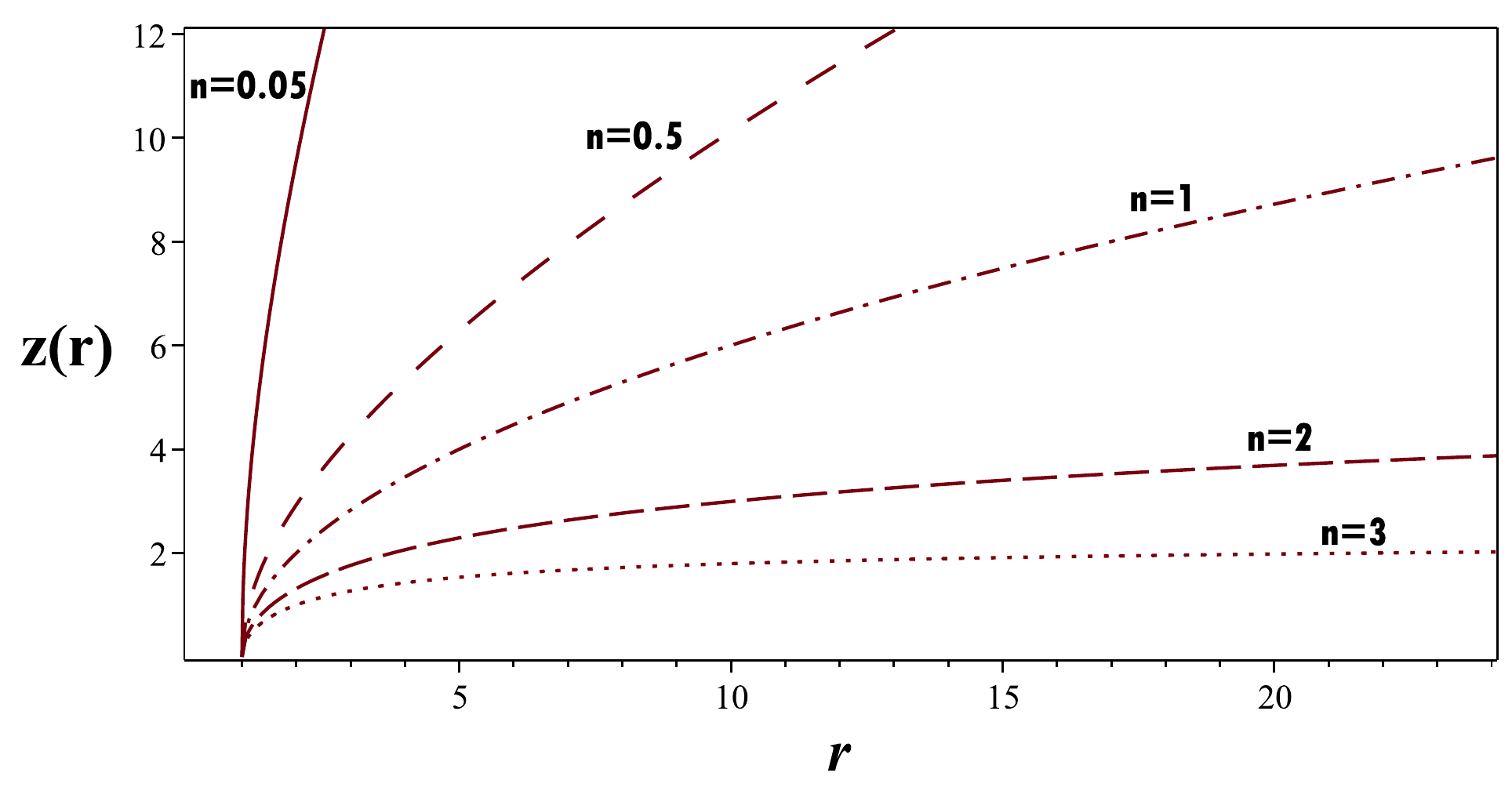}
  \caption{}
  \label{shapen}
\end{subfigure}
\begin{subfigure}{.5\textwidth}
\centering
  \includegraphics[scale=0.43]{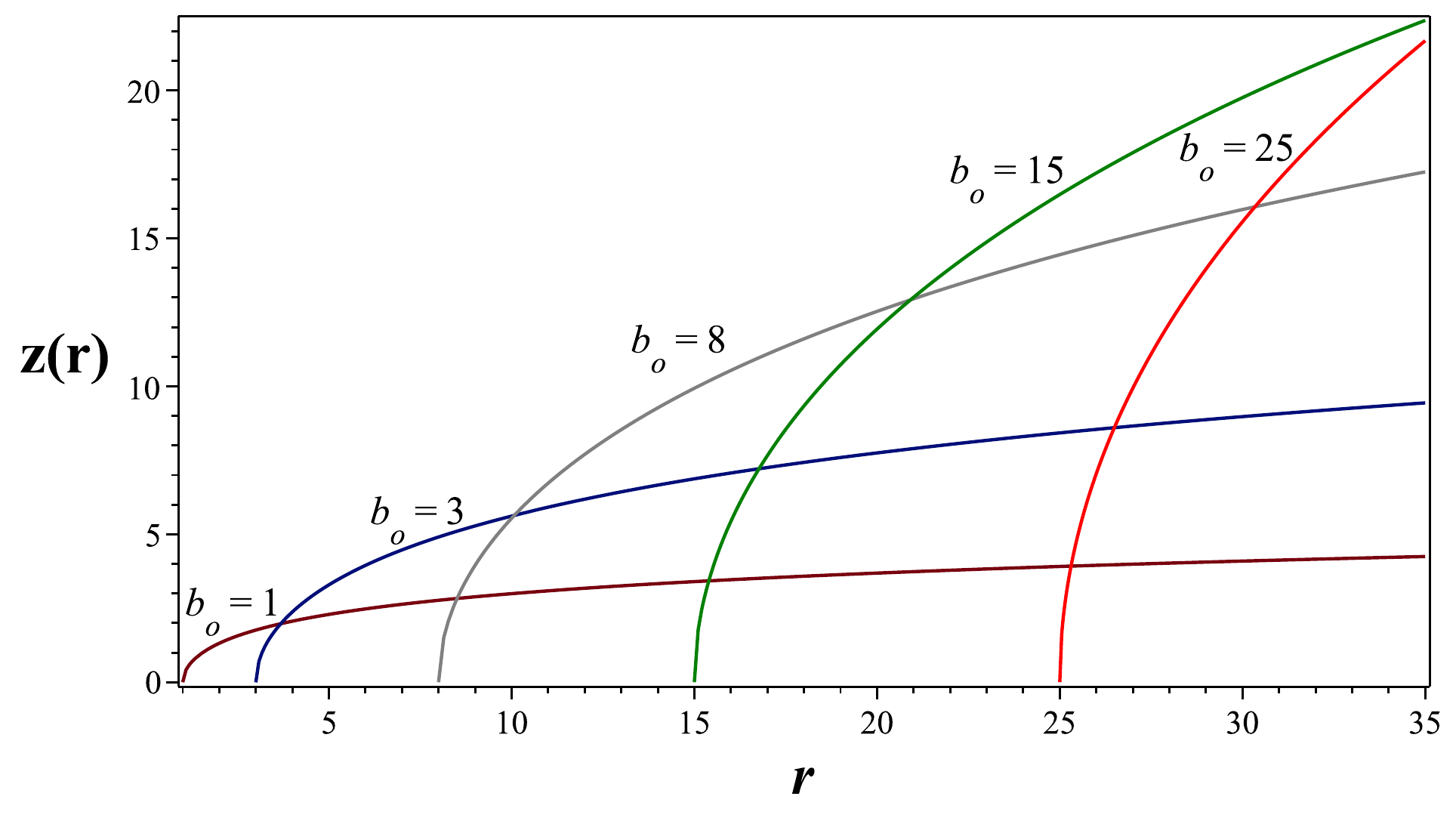}
  \caption{}
  \label{shapebo}
\end{subfigure}
\caption{(a) The plot shows the visualization of wormhole for different values of the shape exponent (n).\hspace{1cm} (b)  The plot shows the visualization of wormhole for different values of the throat radius ($b_o$).}
\label{shape}
\end{figure}

The Morris-Thorne wormhole metric(Ref.\cite{mt}) is given by,
\begin{equation}
ds^2=- e^{2\Phi(r)}dt^2+ \bigg(1-\frac{b(r)}{r} \bigg)^{-1} dr^2 + r^2d\Omega^2 \label{j}
\end{equation}
where $\Phi(r)$ is the redshift function and $b(r)$ is the shape function of the wormhole for which $b(r)\leq r$ and equality holds only at the throat. Both the functions are such that they also satisfy asymptotic flat conditions. Thus, the equation of the trajectory for null geodesics, $Eq.$\eqref{czb}, becomes
\begin{equation} \label{zc}
\boxed{
\frac{1}{r^4}\bigg(\frac{dr}{d\phi} \bigg)^2=\frac{1}{\mu^2} \bigg(1-\frac{b(r)}{r}\bigg)\bigg[e^{-2\Phi(r)}-\frac{\mu^2}{r^2}\bigg] }
\end{equation}
However, note that the coordinate $r$ cannot be used for describing the whole spacetime since it accounts for a coordinate singularity at the throat and is therefore valid for describing geometry only at one side of the throat. Thus, for geodesics that actually reach and pass through the throat, one should not use this formula for the trajectory equation. Instead, one can always work with the proper distance($l$) which must be valid everywhere and throughout the wormhole. As an example, for the metric given in $Eq.$\eqref{a}, we have $dl=\sqrt{B(r)}dr$, thus 
\begin{equation}
l(r)= \pm \int_{b_o}^{r}dr' \sqrt{B(r')} \label{u}
\end{equation}
where, by definition, this proper radial distance is positive for the upper universe, negative for the lower universe and is zero at the throat. Using this, $Eq.$\eqref{czb} can be generalized for wormholes as:
\begin{equation} 
\boxed{
\bigg(\frac{dl}{d\phi} \bigg)^2= \frac{C^2(l)}{\mu^2}\bigg[\frac{1}{A(l)}-\frac{\mu^2}{C(l)}\bigg]}
\end{equation}   
where we have substituted $r$ in terms of $l$ which, in principle, could be obtained by inverting $Eq.$\eqref{u} to get $r\equiv r(l)$. In this paper, however, we will mostly be interested in the behavior of trajectories on one side of a throat and so we will mostly work with $r$ for our convenience.  
Now, using $Eq.$\eqref{i}, the null-geodesics coming from infinity and not reaching the throat gets deflected by an angle, 
\begin{equation}
\boxed{
\alpha(r_o)=-\pi + 2\int_{r_o}^{\infty} \frac{ r_o\ dr }{\sqrt{r[r-b(r)][\exp \{2 \Phi (r_o)-2\Phi(r)\}r^2 - r_o^2 ] } }                
} \label{k}
\end{equation}
It turns out that, for stationary observers in the $r, \theta, \phi$ system, the radial tidal forces can be made to vanish if we have $\Phi'(r)=0$, which we can do by simply choosing $\Phi(r)\equiv 0$, say. This condition gives us a simple class of solutions and corresponds to precisely zero tidal forces. Using $Eq.$\eqref{g}, it can also be deduced that for these wormholes, light can reach the throat only if $|\mu|<b_o$, where $\mu$ is the impact parameter and $b_o$ is the radius of throat. Thus, for these ultra-static wormholes, the light deflection angle becomes,
\begin{equation}
\boxed{
\alpha(r_o)=-\pi + 2\int_{r_o}^{\infty} \frac{ r_o\ dr }{\sqrt{r[r-b(r)][r^2 - r_o^2 ] } }                
} \label{l}
\end{equation}
Now, for an asymptotically flat geometry, a good choice for the function $b(r)$ is, 
\begin{equation}
b(r)=b_o\bigg( \frac{b_o}{r} \bigg)^{n-1}=b_0^nr^{1-n}, \ \ \ n>0 \label{m}
\end{equation} 
where $b_o=b(r_t)=r_t$ corresponds to the throat radius and n=2 gives us the famous Ellis-wormhole\cite{revg}. We will call this parameter `$n$' the \textit{shape exponent.}\\ 
Now, the deflection angle for this choice of $b(r)$ in terms of $r_o$ and $n$ becomes,
\begin{equation}
\boxed{
\alpha(r_o,n)=-\pi + 2\int_{r_o}^{\infty} \frac{ r^{(\frac{n}{2}-1)}r_o\ dr }{\sqrt{(r^n-b_o^n)(r^2 - r_o^2 ) } }                
} \label{s}
\end{equation}
We can see how the deflection angle depends upon the value of shape exponent and the distance of closest approach  as given in Fig. \ref{angle}.
For the Schwarzschild Metric, the deviation angle becomes,
\begin{equation}
\Rightarrow \hspace{3cm} 
\boxed{\alpha(r_o)=-\pi + 2\int_{r_o}^{\infty} \frac{ (r_o/r) dr }{\sqrt{ r^2 \big(1-\frac{2M}{r_o}\big)-r_0^2\big(1-\frac{2M}{r}\big) } } } \hspace{2cm} \label{t}
\end{equation}
It turns out that, for Ellis wormhole, we can write the exact expression for $\alpha$(Ref.\cite{ring}), as
\begin{equation}
\boxed{ \alpha(r_o)=\pi \sum_{n=1}^{\infty}\bigg[\frac{(2n-1)!!}{(2n)!!}\bigg]^2\bigg(\frac{b_o}{r_o}\bigg)^{2n}}
\end{equation}
where we have written $|\mu|=r_o$. In the weak-field regime where $|\mu|<<b_o$, the deflection angle becomes,
\begin{equation}
\alpha(r_o)\approx \frac{\pi}{4}\bigg(\frac{b_o}{r_o}\bigg)^2+\frac{9\pi}{64}\bigg(\frac{b_o}{r_o}\bigg)^4+O\bigg[\bigg(\frac{b_o}{r_o}\bigg)^6 \bigg]
\end{equation}\\
Now, we wish visualize the geometry of such a wormhole, i.e. $Eq.$\eqref{j}, (Ref.\cite{mt}). Since the geometry is spherically symmetric and static, we can confine our attention to an equatorial slice through our wormhole at any instant in time. Then the metric becomes
\begin{equation}
ds^2=\bigg(1-\frac{b(r)}{r} \bigg)^{-1} dr^2 + r^2d\phi^2\hspace{0.8cm} \{\because \theta=\pi/2, t=const.\} \label{n}
\end{equation} 
Now to visualize this 2D geometry, we can embed it into a higher dimensional space of ${\rm I\!R^3}$,i.e., ordinary 3D Euclidean space whose metric can be written in the form:
\begin{equation}
ds^2=dz^2+dr^2+r^2d\phi^2 \label{o}
\end{equation}
Then we can show that,
\begin{equation} \label{p}
\frac{dz}{dr}=\pm \bigg(\frac{r}{b(r)}-1 \bigg)^{-1/2} 
\end{equation}
For a wormhole of type $Eq.$\eqref{m}, it becomes
\begin{equation}
\frac{dz}{dr}=\pm \bigg(\frac{r^n}{b_o^n} -1 \bigg)^{-1/2} \label{q}
\end{equation}
Thus, after integrating, the embedding function becomes(Ref.\cite{taylor}),
\begin{equation}  \label{r}
z(r,n)=\textit{i}\ r\ _2F_1\bigg(\bigg[\frac{1}{2}, \frac{1}{n}\bigg], \bigg[\frac{n+1}{n}\bigg], (r/b_o)^n\bigg)-\textit{i}\ b_o \sqrt{\pi}\frac{\Gamma\big(1+\frac{1}{n} \big)}{\Gamma\big(\frac{1}{2}+\frac{1}{n} \big)}
\end{equation}
where $_2F_1$ represents the hypergeometric function. Note that due to spherical symmetry, there is no dependence on the coordinate $\phi$. Thus, we can easily visualize the geometry through a 2D plot as in Fig. \ref{shape} which represents the embedding function for various values of the shape
exponent and throat radius.
\\
We can also study how proper length $l$ depends upon the radial coordinate $r$ and the shape exponent \textit{n}. If we keep $b_o=1$, then we have
\begin{equation}
l(r,n)=\pm \int_{1}^{r} \bigg(1-\frac{1}{{r'}^n}\bigg)^{-\frac{1}{2}} dr' \label{v}
\end{equation}
\begin{equation}
\Rightarrow \hspace{1.2cm} l(r,n)=\pm r\ _2F_1\bigg(0.5,-\frac{1}{n};1-\frac{1}{n};r^{-n}\bigg) \mp _2F_1\bigg(0.5,-\frac{1}{n};1-\frac{1}{n};1\bigg) 
\end{equation}

\begin{figure}[t] 
\begin{subfigure}{.5\textwidth}
\centering
  \includegraphics[scale=0.6]{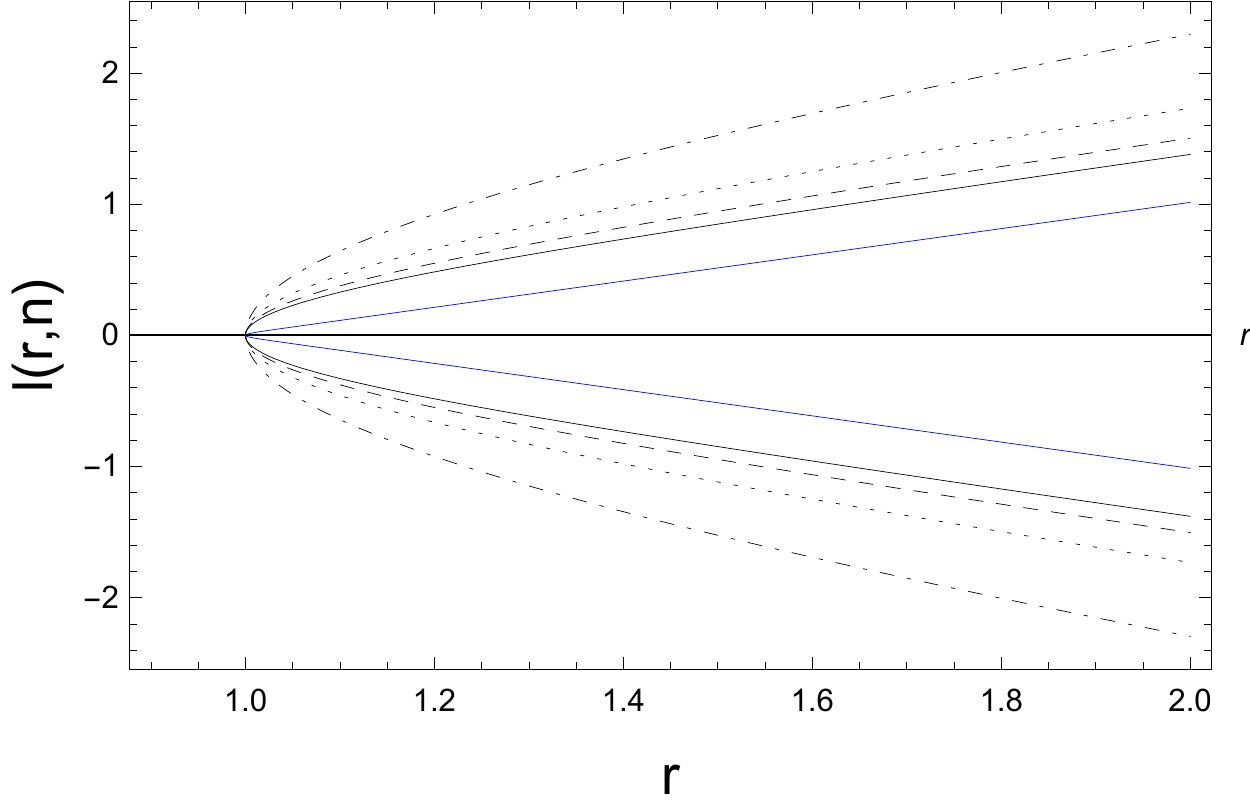}
  \caption{ The plot shows how proper length depends upon the radial coordinate $r$ and the shape exponent $n$.}
  \label{lnr1}
\end{subfigure}
\begin{subfigure}{.5\textwidth}
\centering
  \includegraphics[scale=0.87]{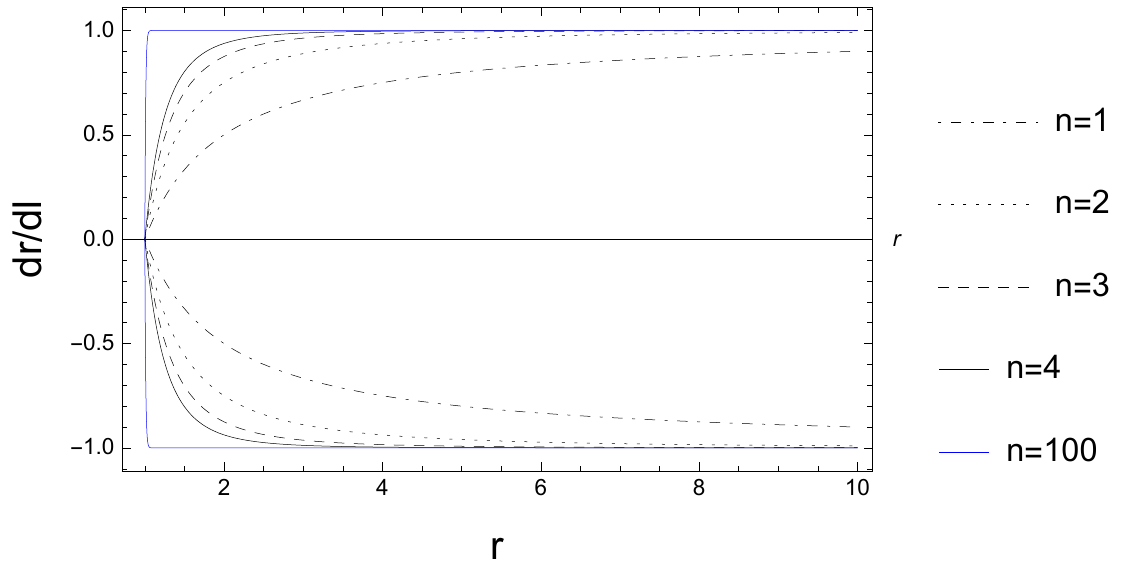}
  \caption{ The plot shows how $dr/dl$ varies with $r$.}
  \label{lnr2}
\end{subfigure}
\caption{In both plots, it can be seen that with increasing $r$ the slope tends asymptotically to $1$, thus corresponding to an asymptotically flat geometry. Also, the slope approaches 1 even faster when the value of $n$ increases. Note that the throat radius has been set to $b_o = 1$.} 
\label{fc}
\end{figure}

A comparision of proper lengths for different geometries is shown in Fig.\ref{fc}.

\vspace{5pt}

\subsection{Photon spheres} \label{psp}
A photon sphere is a location where the curvature of spacetime is such that even null geodesics can travel in circles. In other words, a photon sphere is a region where both $\dot{l}$ and $\ddot{l}$ vanish for a photon. We will calculate the possibility for such a region considering a general static metric as given in $Eq.$\eqref{a}. Rewriting $Eq.$\eqref{e} in terms of \textit{l} gives,
\begin{equation}
\dot{l}^2=\frac{E^2}{A(r)}-\frac{L^2}{C(r)} \label{w}
\end{equation}
Thus, a photon sphere will exist at a location where,
\begin{equation} \label{z}
\boxed{ \frac{dC/dl}{C}=\frac{dA/dl}{A}}
\end{equation}
Another way of looking at it is that, it is the region for which the deviation angle of a photon diverges. Thus, for Morris-Thorne wormhole, \textit{Eq.} \ref{k} implies that the throat itself is a photon sphere which is evident from Fig. \ref{plotnmt}. Also, by \textit{Eq.} \ref{z}, any other place which satisfies the following condition contains a photon sphere,
\begin{equation}
\boxed{r\Phi'(r)=1} \hspace{0.5cm} \text{\{condition for photon sphere to exist at any } r>b_o\} \label{aa}
\end{equation}
\vspace{2pt}

\subsection{Timelike geodesics}
For timelike-geodesics, we have $p_\mu p^\mu=-m^2$ where m is the mass of the particle. If we define the quantities $\widetilde{E}$ and $\widetilde{L}$ as the energy per unit mass ($E/m$) and the angular momentum per unit mass ($L/m$) respectively, then
\begin{equation}
\boxed{ \dot{r}^2=\frac{1}{B(r)} \bigg( \frac{\widetilde{E}^2}{A(r)}-\frac{\widetilde{L}^2}{C(r)}-1\bigg)} \label{ac}
\end{equation}
Notice that a timelike particle always reaches the throat with zero radial velocity, independent of the value of the impact parameter $\mu\ (\because {B(r)}^{-1}=0$ at throat). 
Now, the general equation of the trajectory becomes
\begin{equation} \label{zb}
\boxed{
\bigg(\frac{dl}{d\phi} \bigg)^2= \frac{C^2(l)}{\mu^2}\bigg[\frac{1}{A(l)}-\frac{\mu^2}{C(l)}-\frac{1}{\widetilde{E}^2}\bigg]} 
\end{equation}
where $\mu=\widetilde{L}/\widetilde{E}$ and $l$ is the proper length.
If we differentiate $Eq.$\eqref{ac} with respect to the affine parameter, we obtain for the second derivative of the radial coordinate
\begin{equation}
\ddot{r}=-\frac{B'(r)}{2B(r)}\dot{r}^2+\frac{1}{2B(r)}\bigg[\frac{\widetilde{L}^2}{{C(r)}^2}C'(r)-\frac{\widetilde{E}^2}{{A(r)}^2}A'(r)\bigg]
\end{equation}
The dependence on $\widetilde{L}^2$ is a consequence of spherical symmetry as it tells that the orientation of the angular momentum does not affect the radial acceleration. For a Morris-Thorne wormhole, it becomes
\begin{equation}
\ddot{r}=\frac{1}{2}\bigg(1-\frac{b(r)}{r}\bigg)^{-1}\bigg[\frac{b(r)-rb'(r)}{r^2} \bigg]\dot{r}^2+\bigg(1-\frac{b(r)}{r} \bigg)\bigg[\frac{\widetilde{L}^2}{r^3}-\frac{\widetilde{E}^2}{e^{2\Phi(r)}}\Phi'(r)\bigg]
\end{equation}
For a particle with zero initial velocity $(\dot{r}=\widetilde{L}=0),$ $\ddot{r}\propto -\Phi'(r)$. Thus in ultra-static wormholes, a particle stays at the same position if not given any initial velocity. Also,  at the throat, $\dot{r}=0$ also implies $\ddot{r}=0$. Thus, a particle reaching throat not only attains a zero radial velocity but also has vanishing radial acceleration. The expression for the radial acceleration in an ultra-static wormhole with shape exponent reduces to:
\begin{equation}
\ddot{r}=\frac{nb_o^n}{2r^{n+1}}\bigg[\widetilde{E}^2-\frac{\widetilde{L}^2}{r^2}-1\bigg]+ \bigg(1-\frac{b_o^n}{r^n}\bigg)\frac{\widetilde{L}^2}{r^3}
\end{equation}
The case n=1 is studied in detail in Ref.\cite{reva}. However, note that if $\widetilde{L}=0$, then $\ddot{r}\equiv 0$ for $\widetilde{E}=1$, while on the other hand, $\ddot{r}>0$ for $\widetilde{E}>1$. Thus, this family of geometries correspond to repulsive gravity.

\subsubsection{Unbounded orbits}
If the particle falling from infinity does not hit the throat, it will get deflected after approaching a closest distance of $r_o$, where $r_o$ is then the real solution of the equation, 
\begin{equation}
\frac{\widetilde{E}^2}{A(r_o)}-\frac{\widetilde{L}^2}{C(r_o)}=1 \label{ad}
\end{equation}
Using $Eq.$\eqref{zb} and $Eq.$\eqref{ad}, we can then write
\begin{equation}
\boxed{
\frac{d\phi}{dr} = \pm  \frac{[\mu/C(r)]\sqrt{B(r)}}{\sqrt{\mu^2\bigg(\frac{1}{C(r_o)}-\frac{1}{C(r)}\bigg)+ \bigg( \frac{1}{A(r)}-\frac{1}{A(r_o)} \bigg)}}
} \label{ae}
\end{equation} 
Now, if the particle does not fall into the throat, the total deflection angle ($\alpha$) for a particle falling from infinity will be,
\begin{equation}
\boxed{\alpha(r_o)=-\pi + 2\bigintss_{r_o}^{\infty} \frac{[\mu/C(r)]\sqrt{B(r)}\ dr}{\sqrt{\mu^2\bigg(\frac{1}{C(r_o)}-\frac{1}{C(r)}\bigg)+ \bigg( \frac{1}{A(r)}-\frac{1}{A(r_o)} \bigg)}} } \label{af}
\end{equation}
\vspace{5pt}
For Morris-Thorne wormhole, it becomes,
\begin{equation}
\alpha(r_o)=-\pi + 2\int_{r_o}^{\infty} \frac{\mu r_o \ dr }{\sqrt{r[r-b(r)][\mu ^2(r^2-r_o^2)+r^2r_o^2(\exp{[-2\Phi(r)]}-\exp{[-2\Phi(r_o)]})]}} \label{ag}
\end{equation}

\subsubsection{Bounded orbits}
For a Morris-Thorne wormhole, $Eq.$\eqref{ac} becomes,
\begin{equation}  \label{ah}
\dot{r}^2=\bigg(1-\frac{b(r)}{r} \bigg)\bigg[\frac{\widetilde{E}^2}{e^{2\Phi}}-\frac{\widetilde{L}^2}{r^2}-1\bigg] 
\end{equation}
For ultra-static wormholes, we can simply write
\begin{equation}
\bigg(\frac{dl}{d\lambda} \bigg)^2=\widetilde{E}^2-\bigg(\frac{\widetilde{L}^2}{r^2}+1 \bigg)=\widetilde{E}^2-V^2(r(l))
\end{equation}
where, $dl$ is the differential proper length and $V^2(l)$ can be thought of as the effective potential. This case is studied in detail in Ref.\cite{taylor}. However, we will choose a different form of $e^{2\Phi(r)}$ and will try to study the trajectories it allows. Let us define
\begin{equation}
e^{2\Phi}= \bigg[1-\frac{b(r)}{r}+\epsilon(r) \bigg]\ \label{ai}
\end{equation}
where $\epsilon(r)$ is a continuous function which is significant only near the throat and is vanishingly small otherwise. Now for this choice, we can write a simplified form of $Eq.$\eqref{ah}, for distances far from the throat, thus:
\begin{equation}
\dot{r}^2=\widetilde{E}^2-\bigg(1-\frac{b(r)}{r} \bigg)\bigg( \frac{\widetilde{L}^2}{r^2}+1\bigg) \label{aj}
\end{equation} 
Note that we should not choose $\epsilon(r)\equiv 0$, because then the throat of the wormhole will be a horizon which will make the wormhole non-traversable.\\
Now, we can define an effective potential,
\begin{equation}
\boxed{V^2(r)=\bigg(1-\frac{b(r)}{r} \bigg)\bigg( \frac{\widetilde{L}^2}{r^2}+1\bigg)} \label{ak}
\end{equation}
Therefore, 
\begin{equation}
\boxed{ \dot{r}^2=(\widetilde{E}+V)(\widetilde{E}-V)} \label{al}
\end{equation}
which tells us immediately that the allowed region for a particle with energy $\widetilde{E}$ (as measured at infinity) can be determined from the inequality :
\begin{equation}
\boxed{V(r)< \widetilde{E} }\hspace{1cm} \{ \because\ |V(r)|=V(r) \text{ as } V(r)>0\ \forall\ r>b_o\} \label{am}
\end{equation}
In other words, the radial range of a particle, depending upon its conserved energy $\widetilde{E}$, is bounded within those radii for which $V$ is smaller than $\widetilde{E}$.\\
Also note that, since $r>b_o$, we must have
\begin{equation}
\lim_{r\to\ b_o}V^2(r)=0\ , \  \text {and}\hspace{0.3cm} \lim_{r\to\infty}V^2(r)=1 \label{an}
\end{equation}
It is important to note that any bound orbit that exists around a spherically symmetric source can be of only two types. It can be either a circular orbit (stable or unstable) or an orbit that oscillates around the radius of a stable circular orbit (Ref.\cite{carroll}). So, let us study the possibility of circular orbits in our geometry.\\
\subsubsection*{Circular orbits} 
Now for circular orbits, we require that both $\dot{r}$ and $\ddot{r}$ vanish for at least some r. Therefore,
\begin{align*}
\textbf{Condition I :}\hspace{1cm} \dot{r}&=0 \Rightarrow \widetilde{E}=|V| \\
\textbf{Condition II :}\hspace{1cm} \ddot{r}&=0 \Rightarrow \frac{d}{dr} V^2(r)=0
\end{align*}
It means, for circular orbits, that the energy of a particle should be an extremum of the effective potential. Precisely, if the conserved energy corresponds to a maximum or a saddle point of the potential, then it will be an unstable orbit, while if it corresponds to a minimum of the potential, it will be a stable orbit. \\
Now, if we choose $b(r)=b_o^n r^{1-n}$ as described in $Eq.$\eqref{m}, we can write
\begin{align*}
0=\frac{d}{dr} V^2(r)=(nb_o^n r^{-n-1})\bigg( \frac{\widetilde{L}^2}{r^2}+1\bigg) + \bigg(1-\frac{b_o^n}{r^n} \bigg)\bigg(-\frac{2\widetilde{L}^2}{r^3}\bigg) 
\end{align*}
After simplification, we get
\begin{equation}
\boxed{f(r):= r^n-\bigg(\frac{nb_o ^n}{2 \widetilde{L}^2}\bigg)r^2-b_o^n\bigg(\frac{n}{2}+1 \bigg)=0 }\ \label{ao}
\end{equation}
where we have defined, 
\begin{equation}
V^{2}{'(r)}=\frac{f(r)}{r^{n+3}} \label{ap}
\end{equation}
Also note that, 
\begin{equation}
f(b_o)=-b_o ^n\bigg[\frac{nr^2}{2L^2} + \frac{n}{2}\bigg] \ < 0 \label{aq}
\end{equation}
Now if $r_c$ is some real root of the $Eq.$\eqref{ao}, then a circular orbit is possible only when
\begin{equation}
\textbf{Condition III :}\hspace{1cm} \boxed{r_c>b_o}\hspace{4cm} \label{ar}
\end{equation}\\
Now, let's study what kind of solutions does $Eq.$\eqref{ao}, i.e. $f(r)=0$, have. First we write, 
\begin{equation}
f'(r)=nr\bigg[r^{n-2}-\frac{b_o^n}{\widetilde{L}^2} \bigg]\ , \hspace{0.5cm} f''(r)=n(n-1)r^{n-2}-\frac{n b_o^n}{\widetilde{L}^2} \label{as}
\end{equation}
Let $r_p$ be the point where the first derivative vanishes. Then,
\begin{align}
f'(r_p)=0\hspace{0.3cm} \Rightarrow\hspace{0.3cm} r_p&=\bigg ( \frac{b_o^n}{\widetilde{L}^2}\bigg)^{\frac{1}{n-2}} \hspace{3cm} \label{za}  \\
\Rightarrow \hspace{0.5cm} f''(r_p)&=\frac{nb_o^n}{\widetilde{L}^2}(n-2)\ \hspace{1.8cm} \label{at}
\end{align}
Now, we will consider three cases:\\[5pt]
\textbf{Case I : $n>2$\\}
For this case, it is easy to see that
\begin{equation}
\lim_{r\to0} f(r) < 0, \hspace{0.5cm} \lim_{r\to\infty} f\bigg(\frac{1}{r}\bigg)f(r)<0\ , \hspace{0.5cm} \& \hspace{0.5cm} f''(r_p)>0 \label{au}
\end{equation}

Thus the behavior of f(r) is such that it will start from a negative value at $r=0$ and will grow further negative with the increase in r until it hits a turning point at $r=r_p$, after which it increases monotonically. Thus, it can be inferred that $f(r)$ will have only one positive real root $r_c$ say. Then it is clear that $f(r)>0\ \forall\ r>r_c$. Thus, from $Eq.$\eqref{aq}, we can say that this root must also satisfy condition III. Since there is only one turning point of $f(r)$, it is obvious from $Eq.$\eqref{an} that this corresponds to the maximum of the potential, in which case it will always lead to an unstable orbit.\\ 
Hence, for $n>2$, there will be only one unstable circular orbit for a particular value of $\widetilde{E}$ and $\widetilde{L}$.
\\[8pt]
\textbf{Case II : $n=2$} \\
For n=2, $f(r)$  becomes
\begin{align*}
f(r)=r^2\bigg(1- \frac{b_o^2}{\widetilde{L}^2}\bigg)-2b_o^2
\end{align*}
If $L>b_o$, then 
\begin{equation}
r_c=\frac{\sqrt{2}b_o}{\sqrt{1-b_o^2/\widetilde{L}^2}} > b_o \label{av}
\end{equation}
As we can see, if the conserved angular momentum $\widetilde{L}$ is larger than the throat radius, then we definitely have one root, $r_c$, which satisfies condition III. By the same argument as above, it is clear that it must correspond to a maximum of the potential which can only lead to an unstable circular orbit.\\
Hence, for n=2, there is a possibility of only one unstable circular orbit depending upon $\widetilde{L}$ and $b_o$ and $\widetilde{E}$.
\\[5pt]
\textbf{Case III : $0<n<2$\\}
For this case, we have
\begin{equation}
\lim_{r\to0} f(r) <0\ , \ \ and\ \ \lim_{r\to \infty} f(r) <0 \hspace{0.3cm} \label{aw}
\end{equation}
Thus, it can be seen that, only when $f(r_p)\geq 0$, we have real roots. Precisely, when $f(r_p)=0$ we have one positive real root while if $f(r_p)\geq 0$ we have two positive real roots. Now, we can write
\begin{align*}
f(r_p)=r_p^{n-2}\bigg[1-\frac{n}{2} \bigg]\bigg[ r_p^2-\widetilde{L}^2\bigg(\frac{2+n}{2-n}\bigg)\bigg]
\end{align*}
Since we want $f(r_p)\geq 0$, we can write
\begin{align*}
&r_p^{n-2}\bigg[1-\frac{n}{2} \bigg]\bigg[ r_p^2-\widetilde{L}^2\bigg(\frac{2+n}{2-n}\bigg)\bigg] \geq 0
\end{align*}
\begin{equation}
\Rightarrow \hspace{2cm}\boxed{ \widetilde{L}\geq b_o\bigg[\frac{2-n}{2+n}\bigg]^{\frac{n-2}{2n}}=\Omega }\hspace{2.8cm} \label{ax}
\end{equation}

\begin{wrapfigure}[37]{r}{.45\textwidth}
\begin{subfigure}{0.5\textwidth}
\centering
  \includegraphics[scale=0.4]{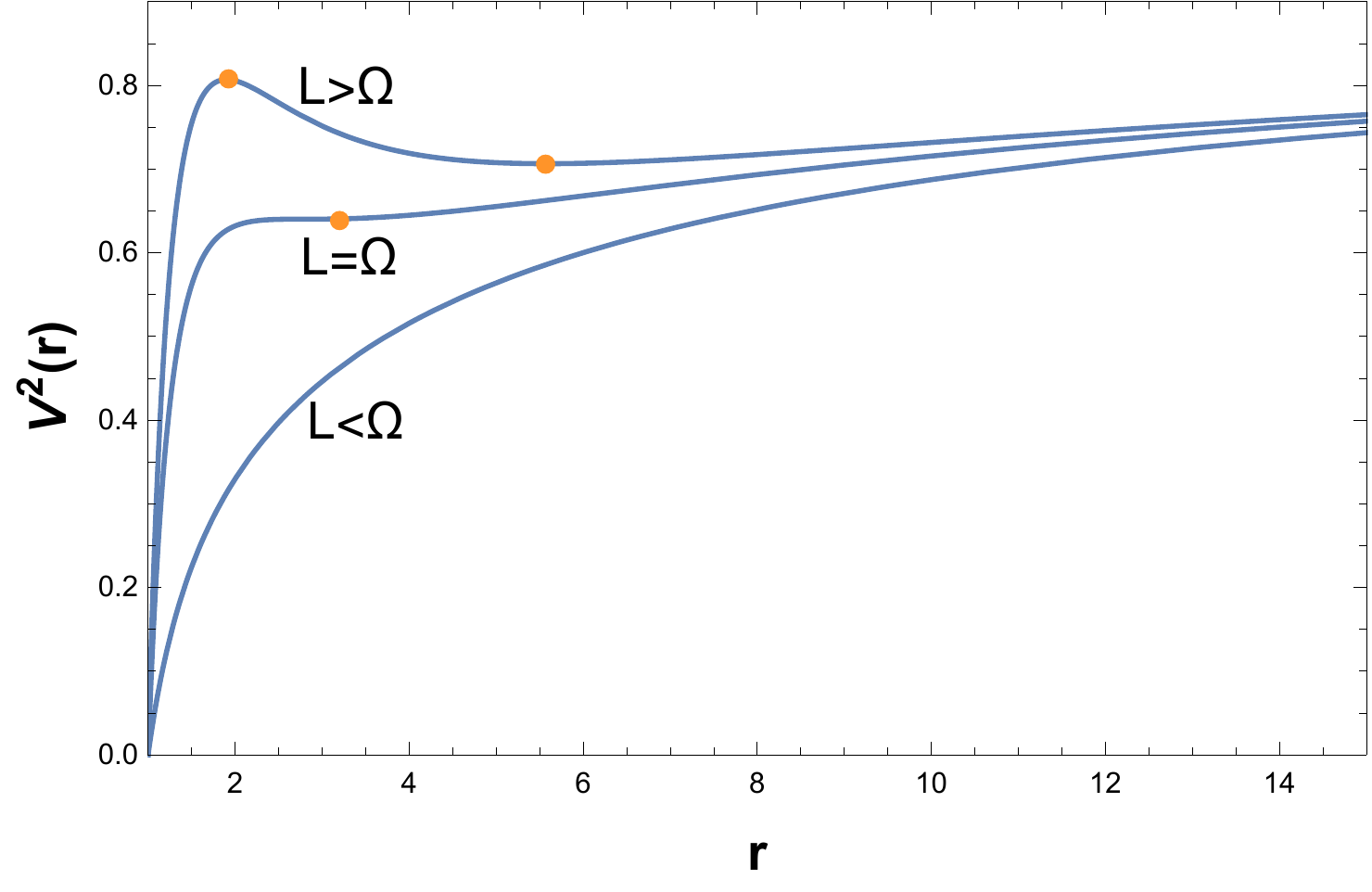}
  \caption{$n<2$}
  \label{plotnl}
\end{subfigure}

\begin{subfigure}{.5\textwidth}
\centering
  \includegraphics[scale=0.45]{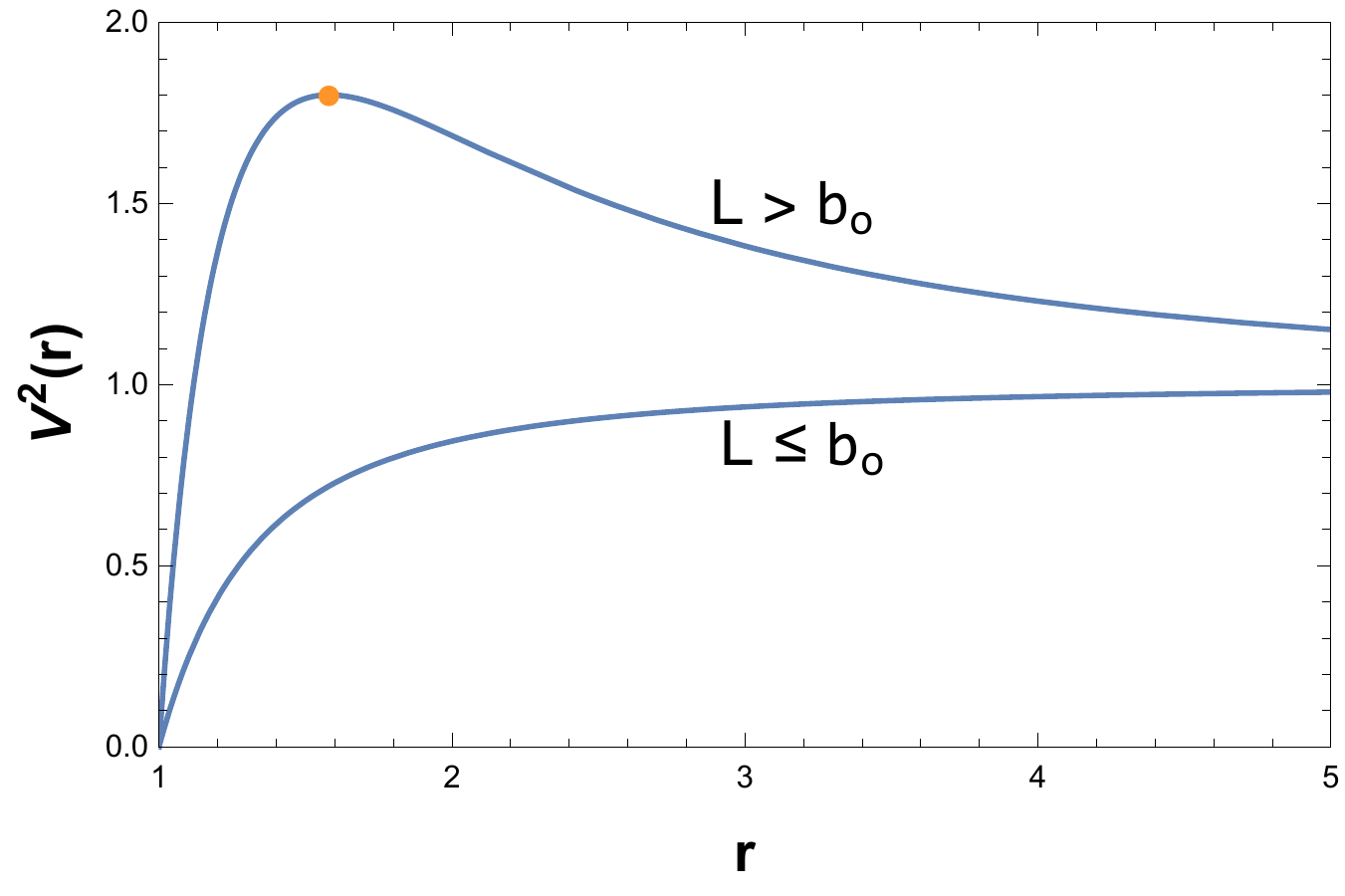}
  \caption{$n=2$}
  \label{plotne}
\end{subfigure}

\begin{subfigure}{.5\textwidth}
\centering
  \includegraphics[scale=0.42]{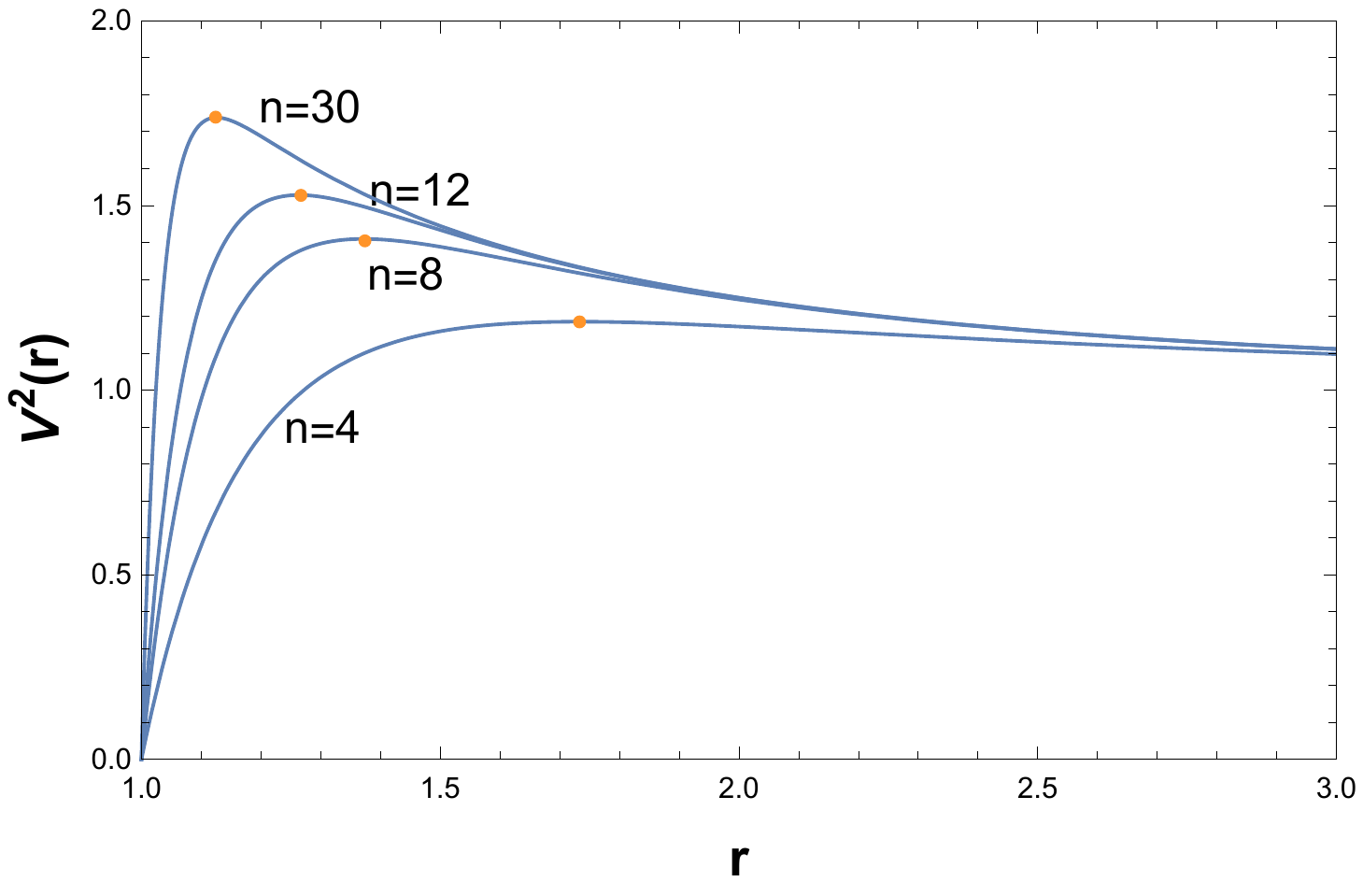}
  \caption{$n>2$}
  \label{plotng}
\end{subfigure}
\caption{Behaviour of $V^{2}(r)\ vs.\ r$ in different cases. Note that this graph does not incorporate the $\epsilon(r)$ contribution near throat and hence is not valid near $b_o$.}
\label{pot}
\end{wrapfigure}

where equality holds for $f(r_p)=0$ and inequality for $f(r_p)>0$.
Note that when $f(r_p)=0$, then, by using $Eq.$\eqref{ap} \& \eqref{za}, we also have $d^2V^2(r_p)/dr^2=0$. Thus, it will correspond to a saddle point of the potential at $r=r_p$. For the condition given in $Eq.$\eqref{ax}, we always have $r_p>b_o$ which means condition III is also satisfied. So, we will have the possibility of one unstable circular orbit for this case.

Meanwhile if $f(r_p) > 0$, we will have two real roots. Now, from $Eq.$\eqref{an}, it is clear that the smaller of these roots will correspond to a local maximum and the larger root will correspond to a local minimum. And, using $Eq.$\eqref{aq} and the condition in $Eq.$\eqref{ax}, it can be inferred that both of these roots will also satisfy cond. III.

Hence, for $0<n<2$, there is a possibility for one unstable circular orbit or a combination of one unstable circular orbit and a stable circular orbit. Note that this is the only case where we have the possibility of stable circular orbits. It is also interesting to note that the Schwarzschild geometry, for which n=1, lies in this case.\\[5pt]
As mentioned before, any bound orbit, which is not circular, is possible only when it oscillates around the radius of a stable circular orbit. Thus, we can say that we surely have no non-circular bound orbits when $ n\geq 2$ for any $\widetilde{E}$ and $\widetilde{L}$ of the particle.\\[3pt]
For the Schwarzschild case (n=1), we can substitute $b_o=2M$ (Schwarzschild radius) with $r$ satisfying $r>b_o$. Then, the condition for circular orbit,$Eq.$\eqref{ax}, becomes
\begin{equation}
\widetilde{L}\geq (2M)\bigg[\frac{2-1}{2+1}\bigg]^{\frac{1-2}{2(1)}}\ \ \Rightarrow \hspace{0.8cm} \boxed {\widetilde{L}\geq \sqrt{12}M } \label{ay}
\end{equation}
which we know is the correct limit for the Schwarzschild case. Thus, what we have done in this section is a general treatment for any shape exponent. But physically, we can say that (Ref.\cite{wald}),
$$V_g(r)=-(b_o/r)^n$$ 
where, $V_g(r)$ is the gravitational potential for a Newtonian like gravitational force given by,
\begin{equation} \label{zd}
F_g(r)=m\ddot{r}=-(nb_o^n)r^{-(n+1)}=-kr^{-(n+1)} 
\end{equation}
So, all our conclusions are valid for this interesting analogy as well. Hence, we have proved, using GR, that in a universe where ``Newtonian like gravity" dies out as $r^{-3}$ or faster, no stable orbits are possible. In other words, the existence of planets will itself be almost impossible.

\subsubsection*{Time period of circular orbits}
We have seen that there is at least one unstable circular orbit possible for any value of the shape exponent. So now, we will try to calculate the time period of a circular orbit of a particle at a distance $r_c$ in terms of $n$ and $b_o$. 
Using $Eq.$ \eqref{ao} and the fact that $E=V(r)$, we can write
\begin{equation}
\boxed{\frac{\dot{t}}{\dot{\phi}}\approx r_c\sqrt{\frac{2r_c^n}{nb_o^n}}} \hspace{2cm} \label{aab}
\end{equation}
Thus, the total time period of a revolution for a circular orbit becomes,
\begin{equation}
\boxed{
\Delta T \approx 2\pi r_c \bigg( \frac{2r_c ^n}{n b_o ^n}\bigg)^{1/2} } \label{aac}
\end{equation}
It means that the velocity required for a satellite to be set in an orbit of radius $r_c$ around a wormhole is given by
\begin{equation}
\boxed{\vec{v}\approx\bigg( \frac{n b_o ^n}{2r_c ^n}\bigg)^{1/2} \hat{\phi}\ } \label{aad}
\end{equation}
By $Eq.$\eqref{aac}, it is also clear that
\begin{equation}
\hspace{2cm}\boxed{\Delta T^2 \propto \frac{r_c^{n+2}}{b_o^n} } \hspace{2.5cm} \label{aae}
\end{equation}                                                                                                                          
As we can see, the time period is always proportional to the radius of the orbit but is inversely proportional to the throat radius. The latter condition signifies that increasing the throat radius can be thought of as keeping the throat radius fixed but decreasing the radius of the circular orbit itself; in which case it is logical that its time period will decrease.\\ 
Again, we can recognize $Eq.$\eqref{aae} as the generalization of Kepler's third law for an attractive force law given by $Eq.$\eqref{zd}. It can be proved immediately by scaling arguments if we put, say, $r'=\lambda r$ and $t'=\mu t$ in $Eq.$\eqref{zd}, to get $\mu ^2 \propto \lambda ^{n+2}$ (Ref.\cite{vbala}).
So, we can retrieve the Kepler's third law in its original form by putting n=1, so that
\begin{equation}
\boxed{\Delta T^2 \propto r_c^{3}} \label{aag}
\end{equation}
For n=2, the time period becomes:
\begin{equation}
\boxed{ \Delta T \approx \frac{2\pi r_c^2}{b_o} } \label{aah}
\end{equation}

\subsubsection*{\normalsize Choice of $\epsilon(r)$}
As we have mentioned, Fig.\ref{pot} is not valid near the throat as it does not consider the significance of $\epsilon(r)$ in that region. Now, we shall try to guess a physically reasonable form for $\epsilon(r)$. 
First, let us consider the problem of tidal forces. For a spaceship whose one end is at $r=a$ and the other end is at $r=b$, the magnitude of the tidal force experienced by the ship would just be the difference between the forces at r=a and r=b. If the spaceship is far away from the throat where $\epsilon(r)<<1$, then the tidal force can be written in terms of the effective potential as:
\begin{equation}
\tau= \bigg|\ \bigg( \frac{d V^2(r)}{dr} \bigg|_{r=b}-\frac{d V^2(r)}{dr} \bigg|_{r=a}\bigg)\  \bigg|
\end{equation}
\begin{wrapfigure}{r}{.4\linewidth}
\centering
\includegraphics[scale=0.5]{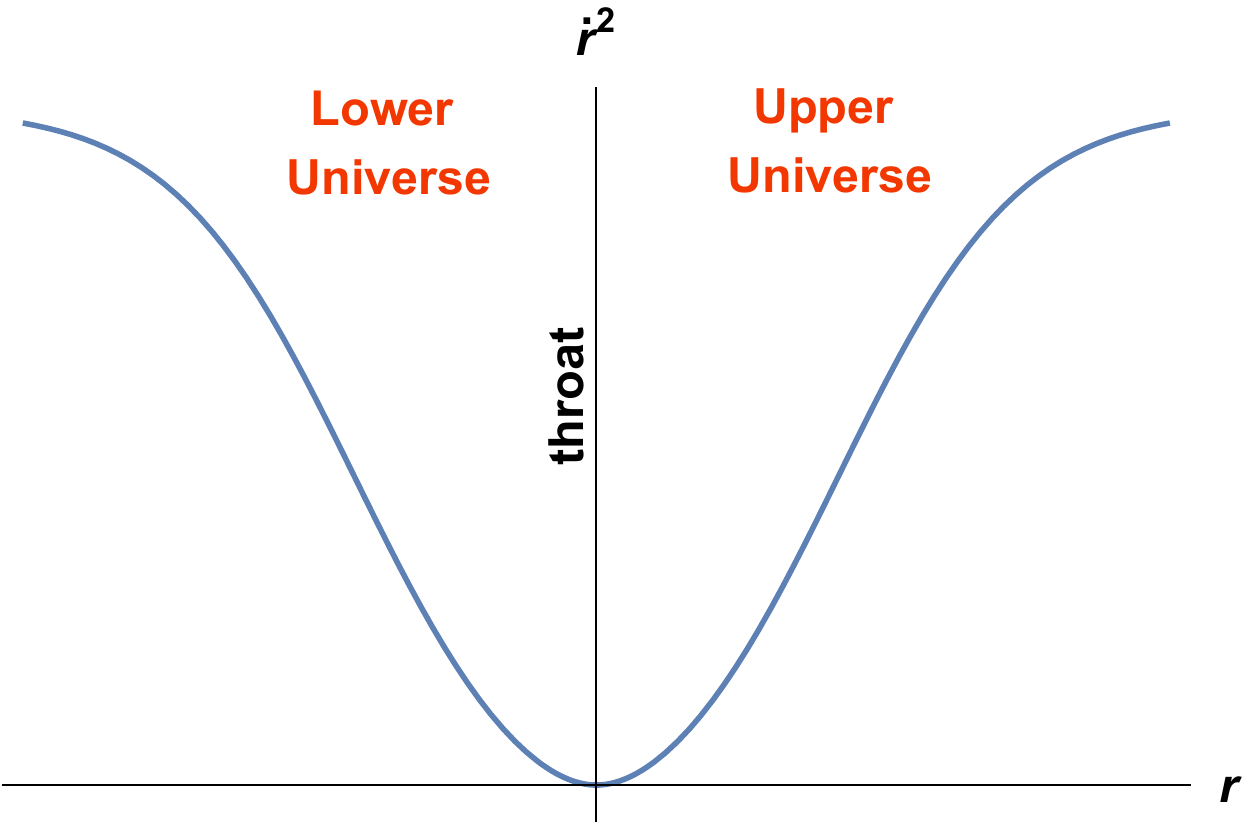}
\caption{A representation of $\dot{r}^2$ vs. $r$ near the throat. The coordinate discontinuity at the throat is ignored.}
\label{thorat}
\end{wrapfigure}
where $\tau$ represents the tidal force. The above expression for the tidal force is obtained by considering the gradient of the potential throughout the spaceship and using the fact that the gradients, except at the two ends of the spaceship, should almost vanish for it to not deviate much from its rigid structure.   Pictorially, it is just the absolute difference between the slopes at the two points on $V^2(r)$ vs. $r$ curve in Fig.\ref{pot}. It can also be noted that without considering the significance of $\epsilon(r)$, we get a discontinuity in the slopes of the potential at the throat. That would correspond to an impulse of force which will be experienced by a particle at the throat while traversing through the wormhole. It would be like hitting a thin
 membrane of a tough material. However, it should be noted that even in the presence of infinite tidal forces, causal contact is never lost among the elements making up the observer; this suggests that curvature divergences may not be as pathological as traditionally thought (Ref.\cite{tidal}).\\
We would definitely want to remove this problem by choosing a reasonable $\epsilon(r)$. If we want a particle to smoothly traverse through the wormhole without any impulse of force, then we can do so by demanding that slope of $V^2(r)$ goes to zero as it reaches the throat. It would imply that\\

\begin{equation}
\lim_{r\to\ b_o} e^{2\Phi(r)} \approx \frac{E^2b_o^2}{L^2+b_o^2}=\epsilon_o \hspace{0.4cm} \{\text{using } Eq.\eqref{ah}\} 
\end{equation}

And since we want it to vanish for large r, a simple choice of $\epsilon(r)$ might be,
\begin{equation}
\epsilon(r)=\epsilon_o e^{1-(r/b_o)^\kappa}, \hspace{0.2cm} \kappa >0
\end{equation}
The larger the $\kappa$, the faster it will die out. We can sketch the plot for $\dot{r}^2$ vs. $r$ as shown in Fig.\ref{thorat}.\\
It is clearly visible that we have removed the discontinuity in the slope at the throat. Also note that there exists a local minimum of $\dot{r}^2$ at the throat. Such a local minimum will correspond to an unstable bound orbit. Thus, for our choice of $\epsilon(r)$, the throat will correspond to a region of unstable circular orbits. 

\vspace{15pt}

\section{Trajectories in a dynamic spherically symmetric wormhole} \label{dynamicwh}
The metric for a spherically symmetric and dynamic Morris-Thorne wormhole can be written as [Ref.\cite{sc}],
\begin{equation}
ds^2=- e^{2\Phi(r)}dt^2+ a^2(t)\Bigg[\bigg(1-\frac{b(r)}{r} \bigg)^{-1} dr^2 + r^2d\Omega^2\Bigg] \label{aai}
\end{equation} 
which corresponds to a 3-geometry with a time dependent scale factor $a(t)$.

\subsection{Null geodesics}
Due to spherical symmetry, we should expect to find the same answer for the deflection angle as that of the static case. For geodesics in equitorial plane, $\theta=\pi/2$ \& $p_\theta=p^\theta=0$. And since $\phi$ is a cyclic coordinate, $p_\phi$ is a constant of motion. So let,
\begin{equation}
p_\phi=L \hspace{3pt}  \Rightarrow \ \ p^\phi=g^{\phi \nu} p_\nu= \frac{L}{a^2r^2}=\dot{\phi} \label{aaj}
\end{equation}
For null geodesics, $ds^2=0$,
\begin{equation}
\Rightarrow \hspace{2cm} \bigg(\frac{ds}{d\lambda}\bigg)^2= - e^{2\Phi(r)}\dot{t}^2+ a^2(t)\bigg(1-\frac{b(r)}{r} \bigg)^{-1}  \dot{r}^2 + a^2(t) r^2  \dot{\phi}^2=0 \hspace{3cm} \label{aak}
\end{equation}
Now, the time-component of the geodesic equation for this metric becomes,
\begin{equation}
\ddot{t}+ \Gamma_{rr}^{t}\dot{r}^2+\Gamma^{t}_{\phi \phi}\dot{\phi}^2+\Gamma^{t}_{rt} \dot{r}\dot{t}=0 \label{aal}
\end{equation}
\begin{equation}
\Rightarrow \hspace{1cm}\ddot{t}+ \frac{a\dot{a}}{\dot{t}} \bigg(1-\frac{b(r)}{r} \bigg)^{-1}e^{-2\Phi}\dot{r}^2+\frac{a\dot{a}}{\dot{t}} r^2e^{-2\Phi}\dot{\phi}^2+ \Phi ' \dot{r}\dot{t}=0 \hspace{1cm}\label{aam}
\end{equation}
Now, substituting the value of $\dot{r}^2$ from $Eq.$ \eqref{aak} and simplifying the expression we get,
\begin{equation}
\frac{d}{d\lambda}[\ln\dot{t} + \ln a + \Phi]=0 \label{aan}
\end{equation}
which upon integration gives,
\begin{equation}
\boxed{
\dot{t}=\frac{E}{a(t)} e^{- \Phi(r)} } \label{aao}
\end{equation}
where, $E$ is a positive constant of integration.\\
Substituting this into $Eq.$\eqref{aak}, we get
\begin{equation}
\dot{r}^2= \frac{1}{a^4(t)}\bigg( 1- \frac{b(r)}{r}\bigg)\bigg(E^2-\frac{L^2}{r^2}\bigg) \label{aap}
\end{equation}
And since $\dot\phi=L/a^2r^2$, the equation of trajectory becomes,
\begin{equation}
\boxed{
\frac{1}{r^4}\bigg(\frac{dr}{d\phi} \bigg)^2=\frac{1}{\mu^2} \bigg(1-\frac{b(r)}{r}\bigg)\bigg[1-\frac{\mu^2}{r^2}\bigg] }
\end{equation}
where, $\mu=L/E$. Note that the time-independence of this equation  is just an artifact of our poorly chosen coordinate system. This is because $r$ is itself a comoving coordinate. We should define a new coordinate, $r'(r,t)=a(t).r$, so that any surface $r' = const.$, $t = const.$ is a two-sphere of area $4\pi r'^2$ and circumference $2\pi r'$.  This coordinate $r'$ can then be called as the ‘curvature coordinate’. In this coordinate, equation of trajectory becomes,
\begin{equation} \label{ze}
\boxed{
a^2(t)\bigg(\frac{du}{d\phi} \bigg)^2=\frac{1}{\mu^2} [1-a(t)b(r'/a)u][1-a^2(t)\mu^2u^2] } \hspace{0.5cm} \{\text{where, }u=1/r' ;\ \mu=L/E\} 
\end{equation}
However, the total deflection angle can be calculated using the coordinate $r$ by the following equation,
\begin{equation}
\boxed{
\alpha(r_o)=-\pi + 2\int_{r_o}^{\infty} \frac{ r_o\ dr }{\sqrt{r[r-b(r)][r^2 - r_o^2 ] } }                
} \label{aaq}
\end{equation}
which is same as $Eq.$\eqref{l}. Also, light will reach the throat only if $|\mu|<b_o$ as in the static case. We can make above conclusions due the fact that the geometry, inspite of a time-dependent scale factor, is always spherically symmetric. 

\subsection{Timelike geodesics}
For simplicity, we will work for the ultra-static case, i.e.,  in which $\Phi(r)=0$. Then, for timelike geodesics, we have
\begin{equation}
- \dot{t}^2+ a^2(t)\bigg(1-\frac{b(r)}{r} \bigg)^{-1}  \dot{r}^2 + a^2(t) r^2  \dot{\phi}^2=-1 \label{aar}
\end{equation}
And the time-component of geodesic equation becomes,
\begin{equation}
\ddot{t}+ \frac{a\dot{a}}{\dot{t}} \bigg(1-\frac{b(r)}{r} \bigg)^{-1}\dot{r}^2+\frac{a\dot{a}}{\dot{t}} r^2\dot{\phi}^2=0 \label{aas}
\end{equation}
Now, from the above two equations, we get
\begin{align*}
\frac{ \dot{t} \ddot{t}}{\dot{t}^2-1} + \frac{\dot{a}}{a}=0
\end{align*}
which upon integration gives,
\begin{equation}
\boxed{ \dot{t}^2=1+\frac{E^2}{a^2} } \hspace{1cm} \label{aat}
\end{equation}
where $E$ is a constant of integration. Using it, we get
\begin{equation}
\dot{r}^2=\frac{1}{a^4}\bigg(1-\frac{b(r)}{r} \bigg) (E^2-\frac{L^2}{r^2}) \label{aau}
\end{equation}
Thus, we will get the same equation of motion as $Eq.$\eqref{ze}. And the deflection angle will be same as $Eq.$\eqref{aaq}.

\vspace{3pt}

\section{Trajectories in a rotating wormhole} \label{rotatingwh} \vspace{-0.25cm}
The metric for a rotating wormhole can be written as (Ref.\cite{teo}),
\begin{equation}
ds^2=-N^2dt^2+\bigg(1-\frac{b(r)}{r}\bigg)^{-1}dr^2+r^2K^2[d\theta^2+\sin^2\theta(d\phi-\omega dt)^2] \label{aaac}
\end{equation}
where $N, K, \omega$ and $\mu$ are functions of r and $\theta$, and $\omega(r, \theta)$ may be interpreted as the angular velocity $d\phi/dt$ of a particle that falls freely from infinity to a point $(r,\theta)$. Assume that $K(r,\theta)$ is a positive, non-decreasing function of $r$ that determines the proper radial distance $R$, i.e., $R \equiv rK$. We also require this metric to be asymptotically flat, which implies
\begin{equation}
\lim_{r\to\infty} N(r)=\lim_{r\to\infty}\bigg(1-\frac{b(r)}{r}\bigg)^{-1}=\lim_{r\to\infty} K(r)=1\ ,\hspace{0.5cm} \lim_{r\to\infty} \omega(r)=0  \hspace{0.5cm} \label{aaad} \\
\end{equation}
However, the metric \eqref{aaac} was initially derived for slowly rotating stars \cite{hartle} and hence it implicitly
assumes the absence of effects due to centrifugal forces \cite{azreg}.
Now, at the equatorial plane, the metric becomes,
\begin{equation}
ds^2=-(N^2-r^2K^2\omega^2)dt^2+\bigg(1-\frac{b(r)}{r}\bigg)^{-1}dr^2+r^2K^2 d\phi^2-2r^2K^2\omega d\phi dt \label{aaaf}
\end{equation}
We can compare it with the metric far from a rotating source of mass M and angular momentum S as given by (Ref.\cite{mtw}), 
\begin{equation}
\begin{split}
ds^2=&-\bigg[1-\frac{2M}{r}+O\bigg(\frac{1}{r^3}\bigg)\bigg]dt^2 - \bigg[4\epsilon_{jkl}\frac{S^k x^l}{r^3} +O\bigg(\frac{1}{r^3}\bigg)\bigg]dt dx^j\\ &+ \bigg[\bigg(1+\frac{2M}{r}\bigg)\delta_{jk}+\bigg(\parbox{12.8em}{gravitational radiation terms\\ that die out as O(1/r)}\bigg)\bigg]dx^jdx^k
\end{split} \label{aaah}
\end{equation}
In cylindrical coordinates, $ x^1=r \cos \phi,\ x^2=r \sin \phi,\ x^3=z$. Assuming axial symmetry, only $S^3$ term survives. Let's call it $J$. Then, 
\begin{align*}
4\epsilon_{jkl}\frac{S^k x^l}{r^3}=\frac{4J[x^1dx^2-x^2dx^1]dt}{r^3}=\frac{4J[r^2d\phi]dt}{r^3}
\end{align*}
Thus, in asymptotically flat limit, comparing it with the $g_{t\phi}$ metric term of the \eqref{aaaf}, we get
\begin{equation}
\boxed{
\omega(r)=\frac{2J}{r^3}+O\bigg(\frac{1}{r^4}\bigg)} \label{aaai}
\end{equation}
\\
Now since the metric terms in \eqref{aaaf} are independent of $t$ and $\phi$, the corresponding momenta one forms are conserved. Thus, we can write (Ref.\cite{roma})
\begin{align}
E&=-p_t=A\dot{t}+B\dot{\phi}\\
L&=p_\phi=-B\dot{t}+C\dot{\phi} \label{aaaj}
\end{align}
where, $\ A=(N^2-r^2K^2\omega^2),\ B=r^2K^2\omega,\ C=r^2K^2$.\\[5pt]
Let, $\ \ \ \  \Delta = AC+B^2=(N^2-r^2K^2\omega^2)(r^2K^2)+(r^2K^2\omega)^2=N^2r^2K^2$ \\[5pt]
Thus, the expressions for $E$ and $L$ in terms of $\dot{t}$ and $\dot{\phi}$ becomes
\begin{equation}
\boxed{
\dot{t}=\frac{CE-BL}{\Delta}\ , \hspace{0.5cm}
\dot{\phi}=\frac{BE+AL}{\Delta}} \label{aaak}  \\
\end{equation}
\subsection{Null geodesics} \vspace{-0.2cm}
For null geodesics, using $Eq.$\eqref{aaaf} and $Eq.$\eqref{aaak}, we get
\begin{align*}
\bigg(\frac{ds}{d\lambda}\bigg)^2=0=\frac{-E(CE-BL)+ L(BE+AL)}{\Delta} + \bigg(1-\frac{b(r)}{r}\bigg)^{-1}\dot{r}^2
\end{align*}

\begin{equation}
\Rightarrow \hspace{0.6cm} \dot{r}^2=\frac{1}{\Delta}\bigg(1-\frac{b(r)}{r}\bigg)(CE^2-2BLE-AL^2) \hspace{1.8cm}   \label{aaal}
\end{equation}
It can be rewritten as
\begin{equation}
\boxed{ \dot{r}^2=\frac{C}{\Delta}\bigg(1-\frac{b(r)}{r}\bigg)(E-V_+)(E-V_-)}   \label{aaam} \\[3pt] 
\end{equation}
where $V_\pm$ are the roots of the equation $CE^2-2BLE-AL^2=0$. Thus
\begin{equation}
\Rightarrow \hspace{0.5cm} V_\pm=\frac{BL\pm |L|\sqrt{\Delta}}{C}\\[3pt] \hspace{0.5cm} \label{aaan}
\end{equation}
Now, since $r,K(r),N(r)$ are all non-negative functions, we get
\begin{equation}
\boxed{V_\pm=\omega L\pm\frac{N|L|}{rK} } \label{aaao}
\end{equation}
Now, a photon will make its closest transit from the wormhole at a distance $r_o$ if at that point the condition, $E=V_{\pm}(r_o)$, is satisfied. If there's no such point, then the photon will definitely fall into the throat. Without loss of generality, we can assume  $J>0$, where $J$ is the angular momentum of the rotating wormhole. This assumption also implies that $\omega(r)>0$. Now, we have two possibilities for the conserved angular momentum $(L)$ of the photon: it can be either $L>0$ or $L<0$.  This will determine whether the light ray is traversing along the direction of frame dragging or opposite to it. \\ 
Also, since we have assumed that there are no horizons, the $g_{tt}$ term of the metric can never change sign. Thus,
$$ -g_{tt}=(N^2-r^2K^2\omega^2)>0$$
\begin{equation}
\Rightarrow \hspace{1cm}\boxed{ \omega(r)<  \frac{N(r)}{rK(r)} } \hspace{1cm}  \label{aaap}
\end{equation}  
So, considering the above inequality and using the fact that $E>0$, the condition, $E=V_{\pm}(r_o)$ becomes,
\begin{equation}
E=
\begin{cases}
|L|\big(\omega_o+\frac{N_o}{r_oK_o}\big), & \text{if } L\geq 0\\[0.3cm]
|L|\big(-\omega_o+\frac{N_o}{r_oK_o}\big), & \text{if } L<0
\end{cases} \label{aaaq}
\end{equation}
where, $\omega_o=\omega(r_o),\ N_o=N(r_o)$ and so on. If we denote $r_o '$ the distance of closest approach when $L<0$ and by $r_o$ when $L>0$, then from the above equation, we can write
\begin{equation}
\begin{split}
\frac{r_oK_o}{N_o}=\frac{|L|}{E-|L|\omega_o}\\
\frac{r_o'K_o'}{N_o'}=\frac{|L|}{E+|L|\omega_o}
\end{split} \label{aaar}
\end{equation}
\begin{equation}
\Rightarrow \hspace{0.7cm} \boxed{ \frac{r_oK_o}{N_o}>\frac{r_o'K_o'}{N_o'}}\hspace{1.8cm} \label{aaas}
\end{equation}
If $N(r)$ is a smooth decreasing function, then this equation proves that the distance of closest approach is greater when the light ray is moving in the direction of frame dragging than that of light moving opposite to it. 
Now, from $Eq.$\eqref{aaak} and $Eq.$\eqref{aaal}, we can write the equation of motion of photon trajectory as,
\begin{equation}
\bigg(\frac{dr}{d\phi} \bigg)^2= 
\begin{dcases*}
\frac{\Delta (1-b(r)/r)(CE^2-2BE|L|-AL^2) }{(BE+A|L|)^2}, &if $L\geq 0$\\[0.3cm]
\frac{\Delta (1-b(r)/r)(CE^2+2BE|L|-AL^2) }{(BE-A|L|)^2}, &if $L<0$
\end{dcases*}
\end{equation}
As we can see, the equation of motion of a photon along the direction of frame dragging is different from that of the opposite direction.\\[5pt] 
Now, if a photon does not fall into the throat, it will get deflected according to
\begin{equation}
\bigg(\frac{d\phi}{dr} \bigg)^2=
\begin{dcases*}
\frac{(B+A\mu_{>})^2}{\Delta [1-b(r)/r][C-2B\mu_{>}-A\mu_{>}^2]}, &if $L\geq0$\\[0.3cm]
\frac{(B-A\mu_{<})^2}{\Delta [1-b(r)/r][C+2B\mu_{<}-A\mu_{<}^2]}, &if $L<0$
\end{dcases*} \label{aaau}
\end{equation} 
where,
$$\boxed{\mu_{>}=\frac{|L|}{E}= \bigg(\omega_o+\frac{N_o}{r_oK_o}\bigg)^{-1}} \hspace{0.3cm}  \text{\&} \hspace{0.3cm}  \boxed{\mu_{<}=\frac{|L|}{E}= \bigg(-\omega_o'+\frac{N_o'}{r_o'K_o'}\bigg)^{-1}}$$
According to the above equation, the deflection angle for a photon moving along the direction of the frame dragging will be larger than that of a photon coming the other way \cite{kerr}.   

\subsection{Timelike geodesics}
For timelike geodesics, we have
\begin{equation}
\dot{r}^2=\frac{1}{\Delta}\bigg(1-\frac{b(r)}{r}\bigg)[CE^2-2BLE-(AL^2+\Delta)] \label{aaav}
\end{equation}
where for simplicity, we have denoted $\widetilde{E}$ as simply $E$ and $\widetilde{L}$ as $L$. Now, we can rewrite the above equation as
\begin{equation}
\dot{r}^2=\frac{C}{\Delta}\bigg(1-\frac{b(r)}{r}\bigg)(E-V_{+})(E-V_{-})  \label{aaaw}
\end{equation}
where, 
$$ V_\pm=\frac{BL}{C}\pm \frac{|L|\sqrt{\Delta}}{C}\sqrt{1+\frac{C}{L^2}} \hspace{3.2cm} $$
\begin{equation}
\Rightarrow \hspace{0.5cm} \boxed{ V_\pm=\omega L\pm\frac{N|L|}{rK}\sqrt{1+\frac{r^2K^2}{L^2}}} \hspace{4.2cm} \label{aaax}
\end{equation}
Now, the equation of a particle's trajectory can be written
\begin{equation}
\bigg(\frac{dr}{d\phi} \bigg)^2= 
\begin{dcases*}
\frac{\Delta (1-b(r)/r)[CE^2-2BE|L|-(AL^2+\Delta)] }{(BE+A|L|)^2}, &if $L\geq0$\\[0.3cm]
\frac{\Delta (1-b(r)/r)[CE^2+2BE|L|-(AL^2+\Delta)] }{(BE-A|L|)^2}, &if $L<0$
\end{dcases*}
\end{equation}\\[7pt]
So, for timelike geodesics, if it does not reach the throat, it will be follow a trajectory given by the equation,
\begin{equation}
\bigg(\frac{d\phi}{dr} \bigg)^2=
\begin{dcases*}
\frac{(B+A\mu_{>})^2}{\Delta [1-b(r)/r][C-2B\mu_{>}-(A\mu_{>}^2+\Delta/E^2)]}, &if $L\geq0$\\[0.3cm]
\frac{(B-A\mu_{<})^2}{\Delta [1-b(r)/r][C+2B\mu_{<}-(A\mu_{<}^2+\Delta/E^2)]}, &if $L<0$
\end{dcases*} \label{aaay}
\end{equation}
$$\boxed{\mu_{>}=\frac{|L|}{E}= \bigg[\omega_o+\frac{N_o}{r_oK_o}\sqrt{1+\frac{r_o^2K_o^2}{L^2}}\ \bigg]^{-1}} \hspace{0.3cm}  \text{\&} \hspace{0.3cm}  \boxed{\mu_{<}=\frac{|L|}{E}=  \bigg[-\omega_o'+\frac{N_o'}{r_o'K_o'}\sqrt{1+\frac{r_o'^2K_o'^2}{L^2}}\ \bigg]^{-1}}$$

\section{Invariant angle method of Rindler and Ishak} \label{invariantangle} \vspace{-0.2cm}
In this section, we will calculate the angle between radial and tangential vectors at a point on the photon's trajectory by Invariant angle Method which was proposed by Rindler and Ishak (Ref.\cite{rindler}). Let $\delta $ represent the radial direction and $d$ represent the tangential direction at any point on the photon's trajectory. Let $\psi$ be the angle between them. Then the invariant formula for $\cos\psi$ becomes
\begin{equation}
\cos\psi= \frac{(g_{ij}d^i\delta^j)}{(g_{ij}d^i d^j)^{1/2}(g_{ij}\delta^i\delta^j)^{1/2}} \label{aaw}
\end{equation}
For a photon coming from far left of a source and heading toward the far right while being deflected, the directions $d$ and $\delta$ in the $(r,\phi)$ basis can be written as\\
$d\equiv (\pm dr, -d\phi)=(\mp dr/d\phi, 1)d\phi = (\mp A, 1)d\phi,\hspace{0.5cm} $ where $A =dr/d\phi$\\
$\delta \equiv (dr,0)=(1,0)dr$
\begin{align*}
\Rightarrow \hspace{0.6cm} \cos \psi &=\frac{ g_{rr}d^r\delta^r+g_{\phi\phi}d^\phi\delta^\phi}{(g_{rr}\delta^r\delta^r)^{1/2}(g_{rr}d^r d^r+g_{\phi\phi}d^\phi d^\phi)^{1/2}}=\frac{|A|\sqrt{g_{rr}} }{\sqrt{A^2g_{rr}+g_{\phi \phi}}} \hspace{0.6cm} 
\end{align*}
Rewriting this in the form of $\tan \psi$ (Ref.\cite{pm}), we get
\begin{equation}
\boxed{
\tan\psi= \sqrt{\frac{g_{\phi \phi}}{g_{rr}}}\ \bigg\vert  \frac{d\phi}{dr} \bigg\lvert \ } \\ \\ \label{aax}
\end{equation}
For a general Morris-Thorne wormhole, it becomes
\begin{equation}
\boxed{
\tan\psi=\frac{ r_o }{\sqrt{[\exp \{ \Phi (r_o)-\Phi(r)\}r^2 - r_o^2 ] } }} \label{aaz}
\end{equation}
\\
In the ultra-static limit, it simplifies to
\begin{equation}
\boxed{
\tan\psi=\frac{ r_o}{\sqrt{(r^2 - r_o^2 ) } }} \label{aaaa}
\end{equation}
It is interesting to note that the expression for $\psi$ is independent of the shape function $b(r)$. This independence is true in the case of dynamical and rotating wormhole geometries as well.\\
The expression for $\psi$ in the Schwarzschild geometry takes the form,
\begin{equation}
\boxed{
\tan\psi=\frac{1}{\sqrt{\frac{r^3(r_o-2M)}{r_o^3(r-2M)}-1} }}\\ \label{aaab}
\end{equation}

\section{Short discussion and concluding remarks} \label{conclusion} \vspace{-0.2cm}
A detailed study of particle and photon trajectories has been conducted in the background of wormhole geometry. Starting with the Morris-Thorne wormhole, null geodesics and photon spheres have been analyzed, while for particle trajectories both bounded and unbounded orbits are considered. Subsequently, both null and timelike geodesics are analyzed in the geometry of dynamic spherically symmetric WH and rotating WH. Finally, using the invariant angle method of Rindler and Ishak, the angle between radial and tangential vectors on the photon's trajectory has been evaluated.\\
Based on the above study, we have found that in a Morris-Thorne wormhole and its dynamic and rotating counterparts, the throat itself is a photon sphere. We have also seen that in such geometries, the angle between tangential and radial vectors at any point on a photon’s trajectory is independent of the shape function b(r). The geodesics in ultra-static wormholes with shape exponents have already been studied in great detail in Ref.\cite{taylor}. Also, in Ref.\cite{reva} Cataldo et al. studied a Schwarzschild-like traversable WH which is obtained by putting n=1 with some slight modification. For geodesics, they showed that a test particle which is radially moving towards the throat always reaches it with zero velocity and at a finite time, while for radially outward geodesics the particle velocity tends to a maximum value, reaching infinity. However, in this paper we have shown that it is true for all possible $n$. Also, general conditions for non-radial geodesics were derived which are required to be satisfied in order for it to cross the throat. These results are in agreement with our study and can, roughly, be obtained by putting $n=1$ in our general equations for arbitrary $n$. Similarly, in Ref.\cite{revg}, the Ellis wormhole ($n=2$) is studied in great detail including the behavior of geodesics in such geometry. For  Ellis wormhole, the particles are attracted on one side and are repelled on the other and so the throat is of saddle nature. In our paper, we have mainly stressed on the geodesics that remain on one side of the wormhole, unlike the above mentioned references where geodesics through the throat are studied in detail.\\ Furthermore, we have analyzed the possibility of bounded timelike orbits for different shape exponents in a different WH geometry, which can be regarded as the generalization of the Schwarzschild geometry far from the throat. There, we used the fact that any bounded timelike orbit in a spherically symmetric geometry is either a circular orbit or an orbit that oscillates around the radius of a stable circular orbit. For this geometry, we found that, for a wormhole with shape exponent $n>2$, there always exists the possibility of one unstable circular orbit while for $n=2$, there exists one unstable circular orbit only when $L>b_o$ and no bound orbits otherwise. That means, no non-circular bound orbits exist when shape exponent $n\geq 2$ for any value of the impact parameter. For $0<n<2$, we found that depending upon the value of $L$ it can either have the possibility of one unstable circular orbit or a combination of one unstable circular orbit and a stable circular orbit. While studying trajectories in a rotating wormhole geometry, we have seen that the equations of motion of both photon and particle depend upon whether it is traveling in the direction of frame dragging or opposite to it. \vspace{-0.2cm}
\section*{Acknowledgement}\vspace{-0.25cm}
The authors are thankful to the Inter University Centre for Astronomy and Astrophysics (IUCAA), Pune (India) for their hospitality as the initiation of this work was taken during a visit there. Anuj is also  thankful to the library facility at the department of mathematics of Jadavpur University. \vspace{-0.2cm}

\bibliographystyle{unsrt}
\bibliography{sample}

\end{document}